\pdfoutput=1
\documentclass[preprint,12pt]{elsarticle}

\usepackage{times}

\usepackage{color}
\usepackage{array}
\usepackage{amsbsy,amsmath,amssymb}
\usepackage{graphicx}
\usepackage{subfigure}
\usepackage{longtable}

\newcommand{\revise}[2]{#2}

\begin{document}

\begin{frontmatter}

\title{Model-based assessment of the risks of viral transmission in non-confined crowds}

\affiliation[iLM]{organization={Institut Lumi\`ere Mati\`ere, CNRS \& Universit\'e Claude Bernard Lyon 1},
            city={Villeurbanne},
            postcode={F-69622}, 
            country={France}}
            
\affiliation[IMAG]{organization={Institut Montpelli\'erain Alexander Grothendieck, CNRS, University of Montpellier},
            city={Montpellier},
            postcode={F-34095}, 
            country={France}}

\affiliation[ENTPE]{organization={\'Ecole nationale des travaux publics de l'\'Etat (ENTPE), Universit\'e
de Lyon},
            city={Vaulx-en-Velin},
            postcode={F-69518}, 
            country={France}}        
            
\author[iLM]{Willy Garcia}
\author[IMAG]{Simon Mendez}
\author[iLM,ENTPE]{Baptiste Fray}
\author[iLM]{Alexandre Nicolas
\footnote{Corresponding author. Email: alexandre.nicolas@polytechnique.edu}}

\begin{keyword}
Covid-19 \sep Viral transmission \sep Pedestrian crowds \sep Epidemiology
\end{keyword}


\begin{abstract}
This work assesses the risks of Covid-19 spread in diverse daily-life situations involving crowds of maskless pedestrians, mostly outdoors. More concretely, we develop a method to infer the global number of new infections from patchy observations, by coupling ad hoc spatial models for disease transmission
via respiratory droplets to detailed field-data about pedestrian trajectories and orientations. This allows us to rank the investigated situations by the infection risks that they present; importantly, the obtained hierarchy of risks is
very largely conserved across transmission models: Street caf\'es present the largest average
rate of new infections caused by an attendant, followed by busy outdoor markets, and then metro and train stations, whereas the risks incurred while
walking on fairly busy streets are comparatively quite low. 
While our models only approximate the actual transmission risks, their converging predictions lend credence to these findings. In situations with a moving crowd, density is the main factor influencing the estimated infection rate. Finally, our study explores the  efficiency of street and venue redesigns in mitigating the viral spread: While the benefits of
enforcing one-way foot traffic in (wide) walkways are unclear, changing the geometry of queues substantially affects disease transmission risks.
\end{abstract}

\end{frontmatter}


\section{Introduction}

Efficient collective action to curb the spread of epidemics in general, and of the current Covid-19 pandemic in particular,
 requires input from a variety of disciplinary fields, from microscale fluid dynamics (to
understand the propagation of virus or bacteria-laden droplets \cite{mittal2020flow,bourouiba2020fluid})
to macroscale epidemiology. At present, the weak link between these
two scales hinders the prediction of how the SARS-CoV-2 virus at the
origin of the pandemic 
\revise{will spread}
{spreads}
in a given crowd. 

Gatherings of people are encountered both in enclosed spaces (such as restaurants,
offices, private accommodation or fitness centers), where statistical
data may be insightful \emph{a posteriori} from an epidemiological standpoint \cite{chang2020mobility,hu2020risk,leclerc2020settings,galmiche2021etude}, as well as in non-confined environments.
Most Covid-19 oubreaks are certainly associated with indoor settings
\cite{bulfone2021outdoor}, but nonetheless a minority of infections
-- at least a few percent, as a tentative estimate \cite{shen2020community,leclerc2020settings,weed2020rapid,galmiche2021etude} --
reportedly originate outdoors (e.g., in beer gardens \cite{express2020beer} and other outdoor gatherings \cite{bulfone2021outdoor}, during casual friendly encounters \cite{galmiche2021etude}, or while jogging \cite{guardian2020coronavirus}) or in mixed indoor/outdoor settings
(e.g., on building sites \cite{EuropeanCDC2020August}). Despite their secondary role, viral transmissions 
amidst outdoor crowds may change the fate of an epidemic when the effective reproduction number $R_\mathrm{eff}$ is close to unity.
Moreover, they raise a specific
challenge because they are inherently hard to trace and document,
but also hard to circumscribe, as they bring together unrelated people\footnote{As of March 2021, more than on third of new cases in France were wholly unaware of how they were infected \cite{galmiche2021etude}.}. These difficulties are a hurdle to the control of outbreaks.

Accordingly, recommendations to wear a face covering
outside have been issued far and wide. Some cities in China, France, part of Germany, Italy, Poland, Portugal, Singapore, South Korea,
Spain, some Swiss cantons, and a number of US states, among others, have put in place mask
mandates for some, or all, outdoor activities. Let us say from
the outset that mask-related policies may have a broader impact than
their chief purpose of physically warding off infections \cite{leung2020respiratory,bahl2020face,SAGE_Britain}:
Widespread usage of face coverings attracts every one's attention to
the health situation and may thus promote stronger vigilance and abiding
by other sanitary measures. Furthermore, these policies are constrained
by the legal possibilities in place in each country, the indirect
consequences of the measures, public perception, and an imperative
of simplicity. On the downside, there is some discomfort associated
with wearing a mask and it might still be too early to assess the psychological impact of being surrounded by covered faces.

Thus, a proper assessment of the risks of infections incurred by maskless crowds in diverse non-confined spaces is much needed, especially at a time when mask mandates are about to be, or have been, lifted. Not only can
it provide more objective foundations to public policies, but it is instrumental in better targeting the situations where risks are
highest and masks are most crucial, thus justifying stricter
local control, and determining if (and where) it makes sense to restrict
pedestrian mobility on streets and sidewalks. It will also be central to the safe revival of outdoor mass events, which are currently suspended in various countries, notably European ones. 

For the time being, there remains a gap between the thriving experimental
and computational studies focused on measuring the emission and propagation
of respiratory droplets \cite{morawska2009size,bourouiba2014violent,asadi2019aerosol,chen2020short,abkarian2020speech,bourouiba2020fluid,bao2020transmission,li2020mask,feng2020influence}
and the investigations of disease spread at larger scales \cite{ferguson2020report,ferretti2020quantifying,harweg2020agent}.
The former have shed light on the complex, turbulent dynamics that
take place inside the exhaled puff and called into question both the
binary distinction between falling droplets and tinier airborne aerosols \cite{bourouiba2020fluid,chong2021extended},
and the scientific basis of the 2-meter social distancing rule \cite{morawska2020time,Jonesm3223,yang2020towards,chong2021extended}.
However, the translation of these results into 
epidemiological predictions relevant for policy-making \cite{vuorinen2020modelling,poydenot2021risk} is uneasy, and trailing. Poles apart from these microscopic
studies, risk assessments at the scale of a facility or venue by means
of agent-based pedestrian simulations \cite{xiao2020modeling,harweg2020agent,romero2020covid}
or large-scale experiments \cite{moritz2020risk} resort to particularly
crude assumptions with regard to viral transmission. Often, an
individual is considered exposed to the disease when he or she comes
in a given radius (e.g., 2 meters) around an infected person, regardless
of their orientations, overlooking that their head orientations control the direction in which
respiratory droplets are expelled. Moreover, pedestrian dynamics models
are hardly designed to reproduce fine observables such as
precise inter-pedestrian spacings with any reliability, neither in usual times nor in times of pandemic, when pedestrian behaviors
and trajectories are altered to reduce infection risks \cite{pouw2020monitoring,ronchi2020risk}.

To overcome these strong limitations, we collected detailed field
data\footnote{The processed field data are freely available on the Zenodo public repository, with the DOI: 10.5281/zenodo.4527462. The Python scripts used to analyze the data can be obtained by
request to the authors.}, during the pandemic in France, about pedestrian separations and orientations in diverse situations
(hereafter called scenarios), either outdoors or in large, ventilated
indoor facilities, and we developed
a mathematically sound method to infer the rate of new infections
in each scenario from our partial observations. The method rests on
simple \emph{ad hoc} models for viral transmissions, which we introduce
and fit to droplet emission data and existing exposure
studies. While 
\revise{none of these simple models can claim to be accurate}
{these models are individually only coarse approximations of the reality},
they all converge towards a fairly robust ranking of the scenarios
in terms of infection risks. The proposed framework is also useful
to quantify the effect of enhanced physical distancing and to assess the
mitigation efficiency of redesigning certain premises (one-way footpaths, queues, etc.).


\section{Methods}

While one can rely on estimated numbers of contacts between people
to model the spread of an epidemic at regional or national scales
\cite{ferguson2020report,ferretti2020quantifying}, more detailed information about
the viral transmission route and the interactions between people is required
to gauge how a virus will propagate in a given crowd. 
Although small respiratory droplets may evaporate into airborne residues that can accumulate in the air and potentially travel long distances, short-range $(\sim 1\,\mathrm{m})$ transmission is widely believed to prevail
 in non-confined environments,
at least for influenza and the coronavirus \cite{freeman2020covid,chen2020short,bao2020transmission}. These droplets are exhaled when breathing, talking, shouting, panting, coughing or sneezing, mostly through the mouth but also through the nose \cite{asadi2019aerosol,morawska2009size};
the focus must thus be put on their transport.

\subsection{Modeling viral transmission via respiratory
droplets}
\hfill\\

In principle, the instantaneous transmission rate due to droplets emitted at $t_e$ by a contagious individual $E$ and inhaled at $t_r>t_e$ by a `receiver' $R$ reads
\begin{equation}
\nu(t_e,t_r)=T_0^{-1}\,\tilde{\nu}\Big[r,\theta^{E}(t_e),\theta^{R}(t_r),\,t_r-t_e,\,\mathrm{ambient\ flows},\,\mathrm{activity(t_e)}\Big]
\end{equation}
where the characteristic time for infection $T_0 \propto n_{\mathrm{inf}}/c_v$ 
\revise{}
{at a given distance in front of the index patient}
is related to the specifics of the disease (namely, the viral titer $c_v$ in the respiratory fluid and the minimal infectious dose $n_{\mathrm{inf}}$), whereas the function $\tilde{\nu}$ accounts for the fluid dynamics of droplet emission and transport. Its parameters $r$, $\theta^E$, and $\theta^R$ will be clarified in the following. 
Unfortunately, an accurate derivation of $\tilde{\nu}$ from the fluid dynamics of droplet and aerosol
propagation in these scenarios would not only be extremely demanding
computationally, but also hinge on very specific information that
is neither available \cite{rosti2020fluid} nor transferable between situations, e.g., ambient
air flows \cite{Jonesm3223,bhagat2020effects}, wind speed \cite{feng2020influence},
humidity \cite{chong2021extended}, and speech details \cite{asadi2019aerosol}. 

\begin{figure}[h]
\begin{centering}

\includegraphics[width=0.7\textwidth]{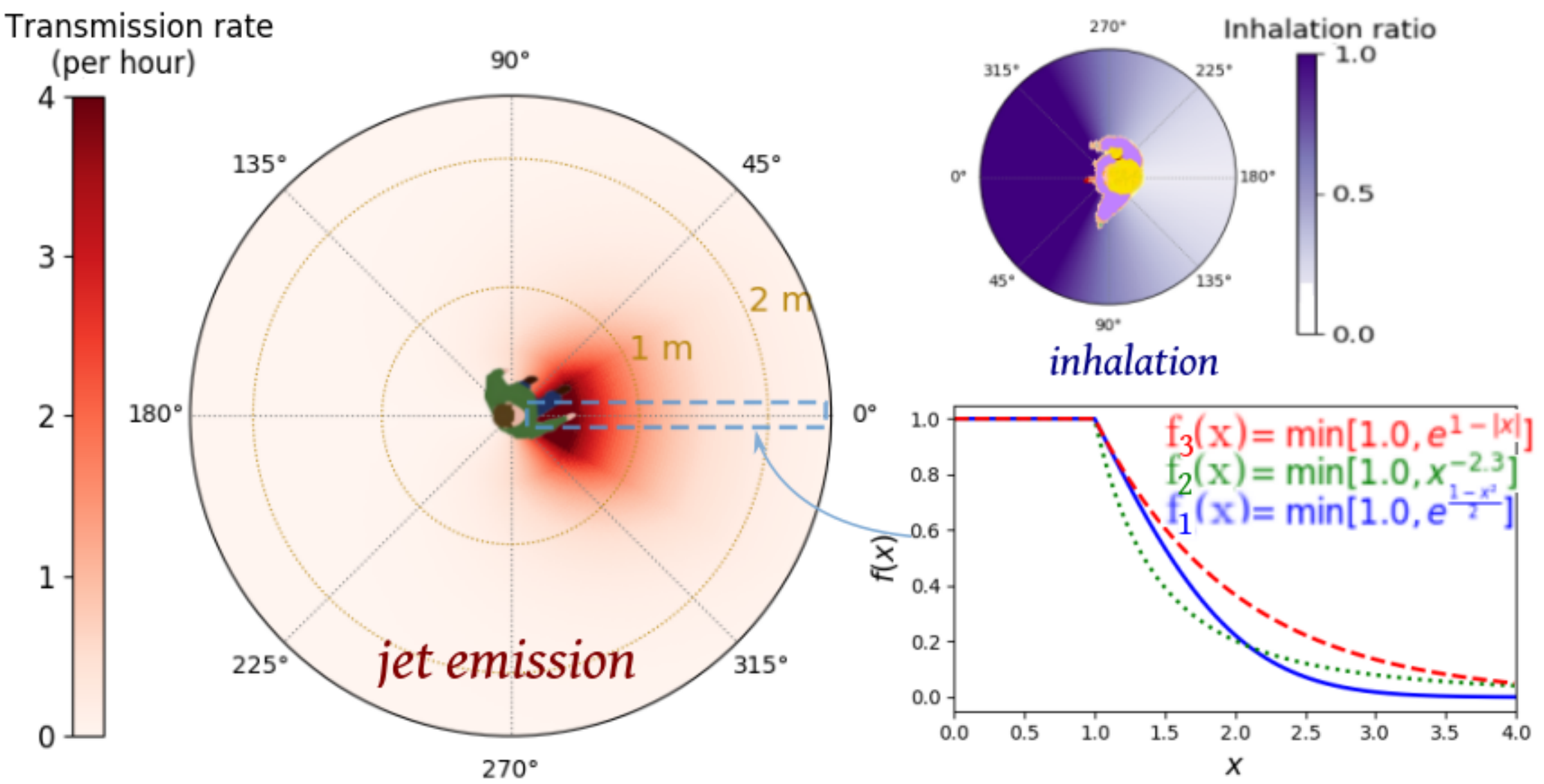}
\par\end{centering}
\caption{ \label{fig:sketch_transmission_rate}
(Color online) Spatially resolved model of disease transmission via virus-laden respiratory droplets. The transmission
rate (Eq.~\ref{eq:nu_expression}) depends on the direction of the emitter's head, the distance
between the emitter and the inhaler, and the latter's head orientation; these dependencies are all modeled with a decaying
function, $f_{1}$, $f_{2}$, or $f_{3}$. Optimistic parameters
combined with $f_{3}$ (\emph{ModOpt}$_3$ \revise{}{model, see Table~\ref{tab:Parameter-sets}}) are used in the illustration. }
\end{figure}

Therefore, we opted for the development
of coarser-grained, \emph{ad hoc} models
that notably overlook propagation delays and ambient air flows. (Relaxing the former hypothesis does not alter our main findings, as we show in Appendix~D by bringing back into play more realistic transmission dynamics \revise{}{(see Fig.~S5)}; on the other hand, ambient air flows may only be neglected if there is very little or no wind, in which case the exhaled puff is expected to be fairly similar indoors and outdoors in the short range.)
With the insight into
droplet propagation gained from computational studies as well
as experiments on expiration and inhalation \cite{abkarian2020speech,li2020assessing,feng2020influence,bourouiba2020fluid,morawska2009size,yang2020towards}, 
the disease transmission rate thus boiled down to a function $\nu_{ER}(t)=\nu\Big[r(t),\theta^{E}(t),\theta^{R}(t)\Big]$ that decreases with the horizontal
distance $r$ between the individuals' heads and the orientations $\theta^{E}$
and $\theta^{R}$ of the emitter's and receiver's heads
\revise{}
{
relative to the direction of the vector that connects them (note that droplet emission and inhalation are not symmetric \cite{abkarian2020speech}).

More precisely, assuming that these variables can
be decoupled, we write the instantaneous transmission rate as

\begin{eqnarray}
\nu(r,\theta^{E},\theta^{R}) & = & \frac{1}{\tilde{T}_{0}}\bar{f}\left(\frac{r}{r_{0}}\right)\cdot\bar{f}\left(\frac{\theta^{E}}{\theta_{0}^{E}}\right)\cdot\bar{f}\left(\frac{\theta^{R}}{\theta_{0}^{R}}\right),\label{eq:nu_expression}
\end{eqnarray}
where $\bar{f}(x)$ is a function such that $\bar{f}(x\approx0)=1$
and $\bar{f}(x)$ decays rapidly for $x\gg1$. To be concrete, we
tested the following three functions,

\[
f_{1}(x)=\exp(\frac{1-x^{2}}{2});\,\,f_{2}(x)=|x|^{-m};\,\,f_{3}(x)=\exp(1-|x|)
\]
Because of the limited accuracy of our positional measurements and
the uncertainties about very near field transmission, the
peaks of these functions are leveled at $x\to0$, viz., $\bar{f}_{k}=\min(1,f_{k})$
for $k=1,\,2,\,3$. 
We then defined a family of plausible parameter sets for the $\nu$ functions, namely, $r_{0}$, $\tilde{T}_{0}$, $\theta_{0}^{E}$,
and $\theta_{0}^{R}$ in Eq.~\ref{eq:nu_expression}) as well as $m$ in
$f_{2}$. The parameters were bounded using a broad scope of available empirical data, which suggest
a characteristic distance $r_{0}\leqslant1\,\mathrm{m}$,
an infection time (at $r_{0}=0.5\,\mathrm{m}$) $T_{0}=\tilde{T}_{0}/\bar{f}(\frac{0.5}{r_{0}})$
between a dozen minutes and an hour, and an exponent $m\approx2-2.5$  (more details
are given in Appendix~B), and within these plausible bounds we explored different sets of values,
listed in Table~\ref{tab:Parameter-sets}. The parameter sets thus span the entire spectrum from highly contagious (`pessimistic')
to only mildly contagious (`optimistic').
}

An example of such a function
is depicted in Fig.~\ref{fig:sketch_transmission_rate}. The spatial decay of the transmission
function is such that $\nu(r,\theta^{E},\theta^{R})$ becomes negligible past a few meters, except for the worst-case
model describing uncovered sneezes \cite{bourouiba2014violent}. The typical time for infection
in the event of a close ($r=50\,\mathrm{cm})$ face-to-face contact lies between 5 minutes (an extremely pessimistic estimate) and 30 minutes (except
for sneezes), consistently with the epidemiological literature on SARS-CoV-2,
outbreak reports and exposure case studies \cite{SAGE_Britain,EuropeanCDC2020August, SantePubliqueFrance2020December, burke2020enhanced, heinzerling2020transmission, chu2020physical,yang2020towards}.

\begin{table}
\noindent \begin{centering}
\begin{tabular}{|c|c|c|c|c|}
\hline 
 & $T_{0}$ (min) & $r_{0}$ (m) & $\theta_{0}^{E}$  & $\theta_{0}^{R}$ \tabularnewline
\hline 
\hline 
Optimistic & 30 & 0.3 & 22.5$^{\circ}$ & 45$^{\circ}$\tabularnewline
\hline 
Moderately optimistic & 15 & 0.5 & 30$^{\circ}$ & 60$^{\circ}$\tabularnewline
\hline 
Standard & 10 & 0.5 & 30$^{\circ}$ & 60$^{\circ}$\tabularnewline
\hline 
Pessimistic & 10 & 0.75 & 45$^{\circ}$ & 60$^{\circ}$\tabularnewline
\hline 
Very pessimistic & 7.5 & 1 & 45$^{\circ}$ & 60$^{\circ}$\tabularnewline
\hline 
Extremely pessimistic & 5 & 1 & 45$^{\circ}$ & 90$^{\circ}$\tabularnewline
\hline 
Uncovered sneezes & 1.7 & 1.5 & 22.5$^{\circ}$ & 60$^{\circ}$\tabularnewline
\hline 
\end{tabular}
\par\end{centering}
\caption{\label{tab:Parameter-sets}Parameter sets used in the transmission model of Eq.~\ref{eq:nu_expression}.
The angular values $\theta_{0}^{E}$ and $\theta_{0}^{R}$ correspond
to the half-angles of the emission and inhalation cones in the horizontal
plane. \revise{}{The model denominations have been chosen to avoid putting an emphasis on a particular parameter of the disease transmission process.}}
\end{table}

The resulting models are representative of a disease that may be transmitted
to multiple individuals within an hour in unfavorable cases \cite{shen2020community,li2020evidence}, but generally
requires close and prolonged contacts for transmission: Even in households,
reported attack rates lie in the range 5\%-30\%\cite{bar2020quantitative}, often around 15\% \cite{park2020early,park2020contact,burke2020active}; moreover, casual episodic contacts at work
or in the community, even face to face, do not necessarily trigger
an outbreak \cite{park2020early,burke2020enhanced}. More quantitatively, the spatial decay described by the fairly optimistic models agrees well with the decay of the concentration of droplets emitted by a coughing subject \cite{li2020assessing} (see Fig.~S3). The reliability of the transmission models is further tested by mimicking a journey aboard a Chinese
train (overlooking its confined nature), where transmission risks for passengers
who sat close to an infected traveler
were recently assessed using trip records \cite{hu2020risk}. Overall, the simulated results, detailed in Appendix~B, are compatible with the empirical data; the models featuring
the most \emph{optimistic} parameters display the best concordance. Finally, let us mention that
the order of magnitude of the parameters (especially the most optimistic ones) is consistent with the
putative minimal infectious dose of SARS-CoV-2, estimated to be order 100 
\revise{}
{
particles
}
\cite{basu2020close}, granted that most exhaled droplets contain 0 or 1 viral copy \cite{poon2020soft} (see Appendix~B). 

Although these pieces of evidence favor the optimistic end of the parameter spectrum, our study is conducted with the whole gamut of plausible model parameters. This variety better reflects our current uncertainty with regard to the transmission
of SARS-CoV-2, but also the established inter-individual and inter-case
variability \cite{morawska2009size,asadi2019aerosol}, depending on physiology, talking characteristics \cite{buonanno2020quantitative}, viral mutations, etc.

\subsection{Field measurements and inference of the rate of new infections}
\hfill\\
\revise{}
{
We used a privacy-respective setup 
to film pedestrian flows and crowds in diverse scenarios in a discrete and passive way
with a TomTom Bandit camera covered with a thin plastic layer (to
deteriorate the quality of the image) and installed in a zenithal position. This privacy-respective
setup was approved by the Data Protection Officer of the French National Centre for Scientific Research (CNRS).

After some pre-processing with the FFMpeg software to correct the 'fish-eye' effect, 
downsample the video, and select the area of interest (from $8\,\mathrm{m^2}$ to $100\,\mathrm{m^2}$), the positions and head orientations
of all pedestrians were manually tracked at a rate of 2 frames per second (fps) with the help of a dedicated Python software; linear interpolation then increased the rate to 10 fps.
The estimated experimental error on the positions is typically
below or around 20~cm, while that on the head orientations is lower than
$20^{\circ}$ in most videos.

Because of the limited field of view, some interactions with off-camera people were missed, especially at large separations.
We compensated for this by appraising the fraction of interactions thus lost  as a function of their range, under the assumption of
homogeneous density, and reweighting the detected contacts accordingly. We checked that this rescaling method successfully
restores the genuine contact distribution, up to contact distances close to the size of the field of view (Fig.~S2 in Appendix~A).

Besides, we were able to identify and keep track of groups of pedestrians (i.e., co-walkers who appear to be relatives, co-workers, or friends) by visual inspection.
Overall, close to 5,000 pedestrian trajectories were thus obtained. Pixel coordinates were converted into real-world coordinates with a geometric formula whose parameters are fit thanks to predefined calibration points at pedestrian
height (see Appendix~A). 
}

For each scenario the time and space-resolved pedestrian measurements are then coupled
to the above transmission models. This directly yields the instantaneous rate, abbreviated as $\nu_{ij}(t)$, at which a supposedly infected index patient $i$, that we will call Iago, transmits the disease to other pedestrians \emph{j} around him via virus-laden droplets. 
Under the independent action hypothesis (IAH) \cite{druett1952bacterial,zwart2009experimental}, each inhaled virus is
equally likely to lead to an infection, with no cooperation or antagonism
between viruses. It follows that, over the time interval $[t_{0},t_{0}+\tau_{i}]$
over which Iago was filmed, he infected a number $C_{i}^{(\tau_{i})}$ of other people $j$ (leaving aside his fellow group members $ G_{i}$, whose possible infection is not specifically related to the scenario, except at the caf\'es), given by a Wells-Riley-like equation \cite{sze2010review}, viz.
\begin{equation}
C_{i}^{(\tau_{i})}=\sum_{j\notin G_i}S_{j}^{0}\cdot\left(1-e^{-N_{ij}}\right),\label{eq:C_tau_i}
\end{equation}
where $S_{j}^{0}$ is the probability
that $j$ is susceptible (i.e., \emph{not already} infected) at the
beginning of the observation interval, and $N_{ij}=\int_{t_{0}}^{t_{0}+\tau_{i}}\nu_{ij}(t)\,dt$
is the cumulative transmission risk \cite{tupper2020event}.

\begin{figure*}
\begin{centering}
\includegraphics[width=1\textwidth]{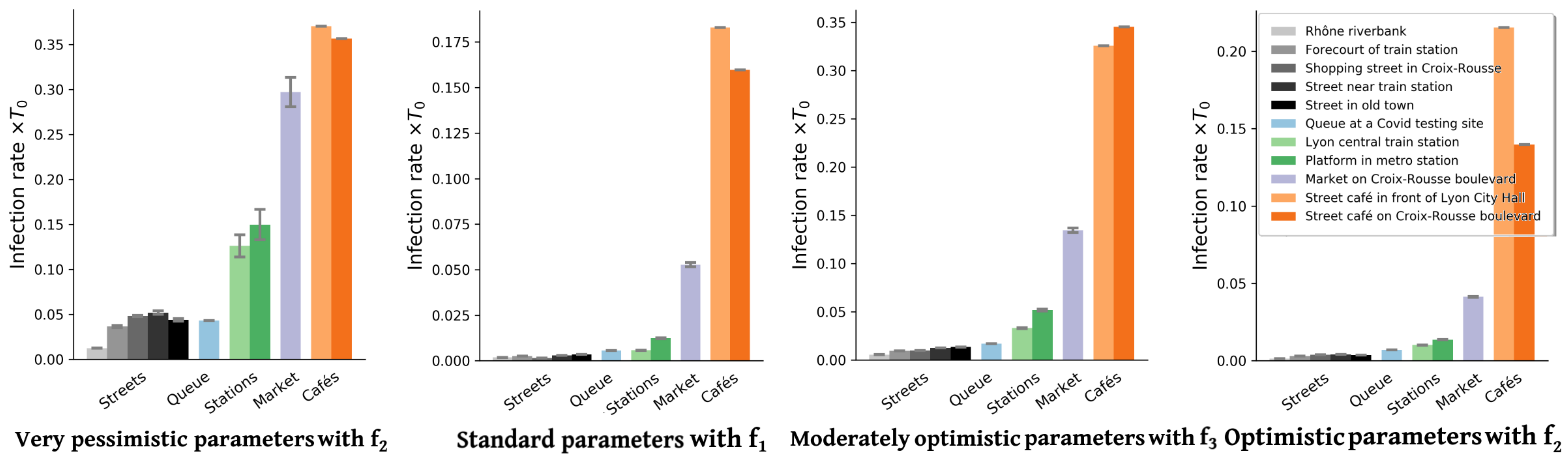}
\par\end{centering}
\caption{\label{fig:Ranking_of_scenarios}
\revise{}
{Number of new
infections over a time interval $T_0$}
in the scenarios under study, estimated with four
different parameter sets for the transmission model. 
\revise{}
{The values taken for the time for infection $T_0$ depend on the model, as detailed in Table~\ref{tab:Parameter-sets};
 in reality, they vary with the index patient, the current activity, the stage of the disease, etc.}
Except in the
static scenarios (caf\'es and queue), infections within groups have
been dismissed and the error bars span the interval between the estimated lower
bound $\underline{C}^{(T_0)}$ and upper bound $\bar{C}^{(T_0)}$,
while the filled bars represent $\frac{1}{2}\left(\underline{C}^{(T_0)}+\bar{C}^{(T_0)}\right)$. Refer to Table~S1 for details about the investigated scenarios.}
\end{figure*}

It is worth noticing that the IAH implies that infections are a stochastic
process without threshold: Any encounter can potentially result
in a new case, and multiple short interactions with various people
are as risky as a single long one, and even riskier because, once
infected, agent \emph{j} can no longer be infected, viz., $S_{j}^{0}:\,1\to0$.
This saturation of the risk complicates the evaluation
of $C_{i}^{(\tau_{i})}$ even if we assume that all pedestrians \emph{except
Iago's group $G_{i}$ of co-walkers} are initially
susceptible, because our videos only record part of Iago's wandering
and may thus miss earlier off-camera interactions. Nonetheless, rigorous upper and lower bounds on $C_{i}^{(\tau_{i})}$ can be set by
noticing that, on the one hand, $S_{j}^{0}\leqslant1$ and that, on the other hand,
$\sum_{j\notin G_{i}}(1-S_{j}^{0})$ cannot be larger than the number
of people actually infected by Iago, which is related to $C_{i}^{(\tau_{i})}$
(see Appendix~C). Finally, for comparison purposes, $C_{i}^{(\tau_{i})}$ is recast
into an hourly rate of new infections $C_{i}\equiv C_{i}^{(\Delta T)}=\frac{\Delta T}{\tau_{i}}C_{i}^{(\tau_{i})}$
with $\Delta T=1\,\mathrm{h}$, assuming that the recorded videos
are representative. As this rate is very sensitive to the chosen characteristic time for infection $T_0$, which exhibits great variability,
we will mostly present results rescaled by $T_0$, which is equivalent to setting $\Delta T=T_0$ and showing the number of new cases over a time
interval $\Delta T$.

Static scenarios -- namely, the caf\'es and waiting
lines -- are handled slightly differently, because then Iago's neighbors
do not change significantly as time passes, in which case 
we set $N_{ij}=\frac{\Delta T}{\tau_{i}}\int_{t_{0}}^{t_{0}+\tau_{i}}\nu_{ij}(t)\,dt$
and $S_{j}^{0}=1$ in Eq.~\ref{eq:C_tau_i}. 
 In a nutshell, the proposed framework
enables us to quantitatively translate patchy observations, with undetected
contacts, into an estimated global risk of viral spread.

\section{Results}

\subsection{Ranking of scenarios by the risks of new infections}
\hfill\\
Inserting the collected field data into this framework,
we obtain upper and lower bounds on the mean
rate $C=\left\langle C_{i}\right\rangle _{i}$
of new infections per hour for each scenario and each transmission model. Figure~\ref{fig:Ranking_of_scenarios} presents a sample of results for four of these models
(also see Figs.~S6-S8 \revise{}{for the results obtained with other models}). These results confirm the efficiency
of the proposed bound-setting method, as the bounds are found to confine $C$ to a narrow interval. Most importantly, the ranking of the different scenarios
turns out to be robust, that is to say, largely preserved across models.
This is our first major result.

Pursuing the analysis of Fig.~\ref{fig:Ranking_of_scenarios}, we observe that street caf\'es
present the highest risks in terms of the mean number of new infections
per hour, even though their tables were more spaced when the videos were shot
than before the pandemic. These infections at caf\'es are easily
rationalized by the close, face-to-face interactions between people sharing
a table, let alone the increased emission of droplets associated
with lively discussions and eating, which is overlooked here. This result
is in line with case reports of high risks of viral transmission while
dining and drinking
(indoors or outdoors, unspecifically) \cite{chang2020mobility,li2020evidence,tupper2020event,galmiche2021etude}. Next in line among the observed scenarios comes the outdoor
market alley. Despite its high average density $\rho\simeq0.5\,\mathrm{ped/m^{2}}$, this scenario never matches the level of risk at caf\'es,
bar with the very pessimistic parameters corresponding to high contagiousness.
Further down the list, crowd density explains the considerably higher risks at
train and metro stations ($\rho\simeq0.25\,\mathrm{ped/m^{2}}$) than on fairly busy streets in Lyon ($\rho\simeq0.05-0.1\,\mathrm{ped/m^{2}}$)
and, to an even larger extent, the riverbank walkway that we filmed.
Somewhat intriguingly, the estimated infection rate may be as
large, or even larger, at the observed testing site in Lyon than it is on
these streets, although the overall density there is low and attendants
were strictly asked to stay 2 meters apart from each other;
yet, their relative proximity was prolonged over considerable time and,
besides, they tended to turn and pace around a bit while waiting.
One should however bear in mind that our models estimate the risks of viral
transmission if no face mask is worn, whereas everybody was wearing
a mask at the testing site that we filmed.

\subsection{Rates of new infections}
\hfill\\
Besides the robustness of this qualitative ranking of scenarios, largely maintained across  parameter sets, on a more quantitative note we observe that the infection rates are always (except with the worst-case models) at least 10 times scantier
in the investigated streets than at caf\'es, even without taking into account that talking and eating augment droplet emissions. In addition, the pessimistic estimates are generally at most a factor 10 larger than the (possibly more relevant) most optimistic ones. Thus, it is reasonable to conclude from those estimated values, that contagious Iago will infect a number of order 1 person \emph{roughly speaking} if he sits at a caf\'e for one hour, whereas he would probably cause \emph{significantly} less than $\sim0.1$ new infections if he spent this time walking on a fairly busy street.

\begin{figure}
\begin{centering}
\includegraphics[width=0.5\textwidth]{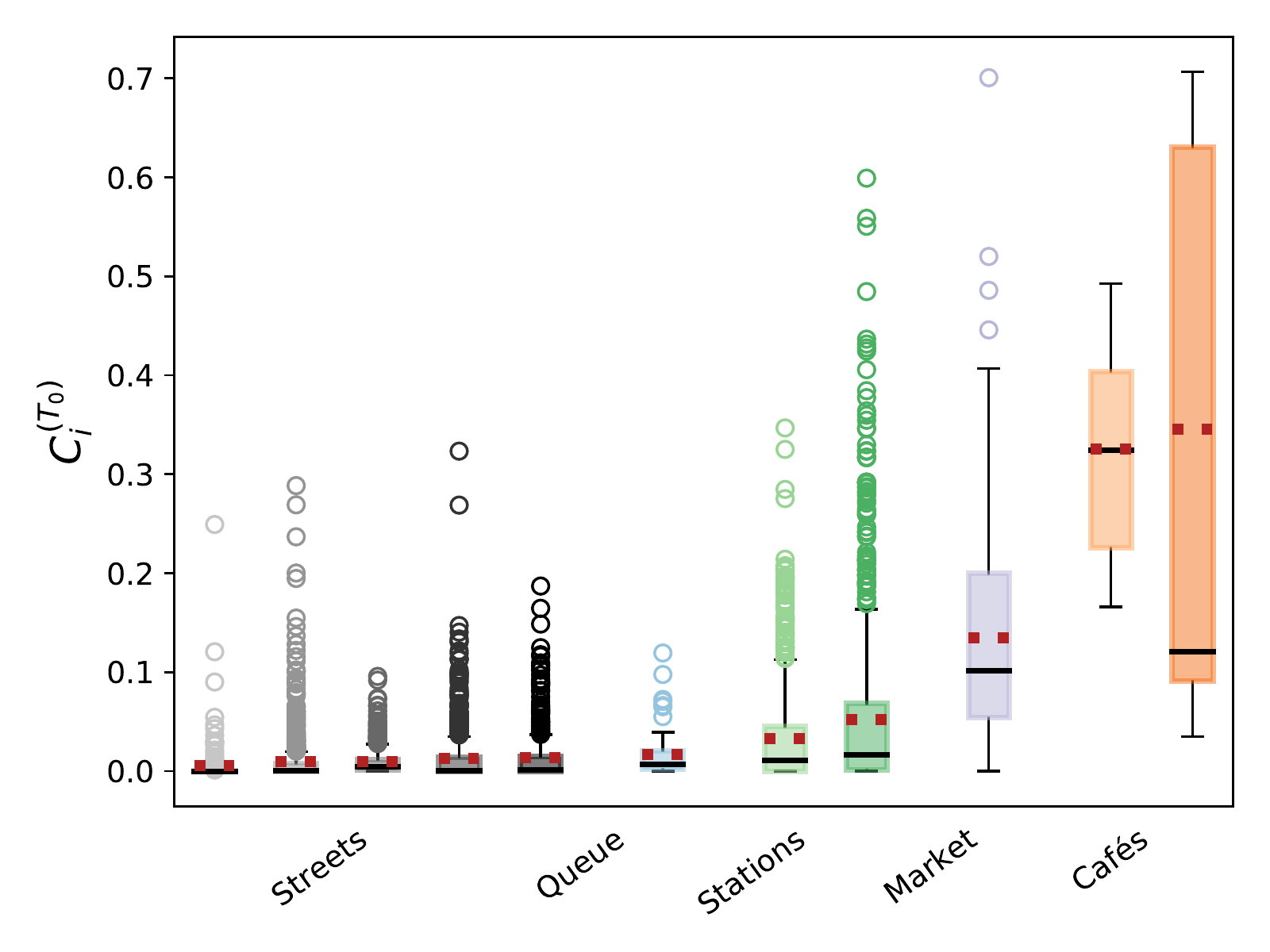}
\par\end{centering}
\caption{\label{fig:boxplot}(Color online)
\revise{}
{Number of new infections $C_i^{(T_0)}$ over a time interval $T_0$}
caused by the different
pedestrians $i$ in each scenario, as estimated with \emph{ModOpt}$_{3}$. 
The dashed red lines represent mean values, solid back lines are medians and open symbols are outliers.
}

\end{figure}

Nonetheless, these average rates brush aside the variety of pedestrian contacts
in the different scenarios, which is better reflected in the box plots
of Fig.~\ref{fig:boxplot}. The figure shows that, while the scenarios
involving a moving crowd cause fewer infections than
caf\'es on average, their rates of infections $C_{i}$ are more dispersed
and, unlike caf\'es, they feature many 
\revise{outliers on both ends, i.e.,}
{
values that significantly deviate from the bulk, both
}
at $C_{i}\simeq0$ and at high-$C_{i}$,
the latter being pedestrians that fortuitously turn into super-spreaders
because of their pattern of on-street contacts. As we shall see below,
the blame does not necessarily rest on the pedestrian, but rather
on the ebbs and flows of crowding in each observed situation.

Prior to that, let us remark that accounting for the directionality
of droplet propagation and describing the orientations of pedestrians'
heads had a marked effect on our results. Indeed, not only does an isotropic
transmission model overestimate risks in crowds by a factor of at least 10 in comparison to its directional counterpart, but it also alters
the ranking of scenarios: It predicts considerably more infections
at the outdoor market than at the caf\'es (Fig.~S8). Otherwise,
such an inversion (along with high risk estimates) is only found for our
worst-case transmission models, in particular the model that we introduced
to mimic the effect of a contagious patient sneezing every few minutes
without covering his or her sneezes. 
On the other hand,
allowing infections within groups, as we did for the caf\'es, does not dramatically change the picture, even though it
substantially heightens the risks associated with sparse situations,
for instance, the riverbank walkway. This is not surprising because
in these situations close contacts mostly occur between group members.

\subsection{Key determinants of the transmission rate}
\begin{figure}
\begin{centering}
\includegraphics[width=0.5\textwidth]{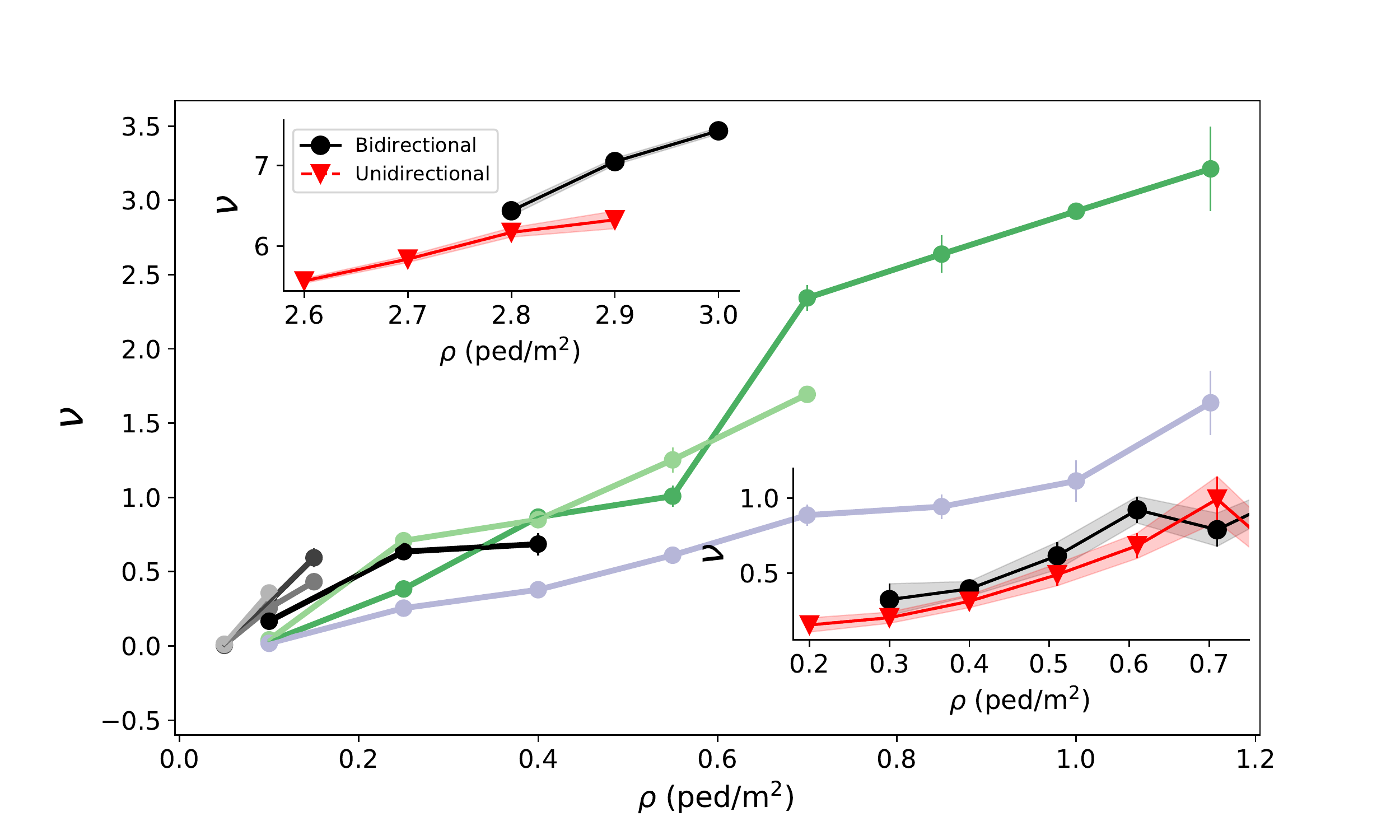}
\par\end{centering}
\caption{\label{fig:risk_vs_density}Dependence of the rate of viral transmission
$\nu(t)$ on the instantaneous density $\rho(t)$, in different scenarios,
using the \emph{ModOpt}$_{3}$ parameter set. Refer to Fig.~\ref{fig:Ranking_of_scenarios}
for the color code. In the \emph{insets}, the rates are distinguished
depending on the directionality of the flow in the (\emph{bottom right}) outdoor market scenario,
(\emph{top left}) corridor flow experiments of the BaSiGo project \cite{cao2017fundamental}. The error bars and envelopes
represent standard errors (i.e., $\pm\frac{\mathrm{std}}{\sqrt{n}}$,
where  $n$ is the number of
uncorrelated data points).}
 
\end{figure}

\hfill\\
To better understand the observed disparities, we need to identify the key
variables that determine the level of risk. Figure~\ref{fig:risk_vs_density}
confirms the intuition that the instantaneous pedestrian density $\rho(t)$
is a major determinant of the rate of viral transmission $\nu(t)\equiv\left\langle \sum_{j\notin G_i}\nu_{ij}(t)\right\rangle _{i}$
(where the average is taken over all pedestrians observed at time
\emph{t}), in that it controls how close each pedestrian is to their counterparts. (Note that all time-dependent variables have been averaged over
intervals of two seconds, to reduce the statistical noise.) The
variation of $\nu$ with $\rho$ looks similar across scenarios, but
is not strictly identical, which indicates that other scenario-dependent
variables affect the transmission rate $\nu$. Furthermore, these
variations become more muddled as one turns to more optimistic parameter
sets, which is consistent with the idea that one then probes the configuration
of the crowd at finer length-scales, owing to the shorter transmission
range. The total pedestrian flow rate could in principle play a role; however, we found that $\nu$ does not follow any
clear trend with this flow rate at fixed density $\rho$ (Fig.~S9). 

On the other hand, head orientations naturally have some bearing on the
risks of infection, as evinced by the failure of isotropic transmission
models to reproduce our results\footnote{We think that this is largely due to the discrepancy between
the face-to-face orientations at caf\'es (which facilitate transmission) and the more or less random orientations e.g. at a market.}, but we now show that in non-static
scenarios these orientations can be practically inferred using only
the trajectories. To do so, we notice that the head orientations of
walking pedestrians (speed $v>0.3\,\mathrm{m\cdot s^{-1}}$) are approximately
normally distributed around their walking direction, with a standard
deviation around $26^{\circ}$, and decorrelate over one second. Therefore,
we choose to ascribe angular orientations randomly drawn from this normal
distribution to walkers, while their stationary counterparts ($v\leqslant0.3\,\mathrm{m\cdot s^{-1}}$)
are considered purely randomly oriented; the random values are refreshed
every second. Quite interestingly, this simple reconstruction of head
orientations yields mean infection rates $C$ per hour that
agree very well with the values
\revise{}
{computed with the \emph{bona fide} orientations,}
with a relative
difference generally lower than 15\%, regardless of the transmission
model. The correspondence between the individual $C_{i}$
values for each pedestrian is of course imperfect with this method,
but overall the differences are not extremely large (Fig.~S10).
These observations are particularly relevant to bolster risk assessments
based on observed or (reliably) simulated pedestrian datasets in which
head orientations are missing, as they most often are.

\section{Discussion}

\subsection{Insight into the risk in the scenarios under study}
\hfill \\
The spatial resolution of our empirical data and
models provides deeper insight into the 
circumstances of infection
in the above scenarios; it can contribute to the debate about what physical
distance should be prescribed between pedestrians in non-confined
environments, and whether 2 meters or '6 feet' are enough from an epidemiological perspective \cite{Jonesm3223,morawska2020time}.
Admittedly, the answer will heavily rest on the transmission model (which was here established in an \emph{ad hoc} way), but the statistics
of inter-pedestrian contacts in the scenario also play a large role.
Using the cumulative rate of transmission $\int\nu(t)\,dt$ as a proxy
for the incurred risk, we find that its dominant contribution comes
from interactions within a distance of 1 meter (for instance, 70\%
at Bellecour metro station, with \emph{ModOpt}$_{3}$), whereas transmission
beyond 2 meters, albeit possible, accounts for only a few percent
of the risk (5\% at Bellecour), at most. 
Importantly, here and throughout the paper, risks have been quantified by the number of new cases expected in each setting; this choice is
relevant at the \emph{collective} scale, for policy-making, but not
for the evaluation of the risks incurred by an \emph{individual} in the crowd.

\subsection{Mitigation efficiency of redesigns}
\hfill \\
\revise{Beyond these conclusions rooted in actual observations,}
{Beyond this debate, }the framework
introduced here opens the door to evaluating the mitigation efficiency
of hypothetical redesigns of streets and venues, consisting e.g. in
enforcing one-way
circulation on footpaths, a sitting plan at caf\'es, or increased spacing in queues. Since
circulation plans have flourished during the pandemic, let us first
explore the impact of one-way \emph{vs.} two-way foot traffic on sidewalks and pedestrian streets.

To avoid potential situational biases, the question is investigated by separating the
periods of time (binned in two-second intervals) when the flow was unidirectional from those when there were pedestrians going in both directions, in each given scenario -- the market alley in the bottom right inset of Fig.~\ref{fig:risk_vs_density}. Since the transmission risks were found to depend on density, but not on the total flow rate (i.e., the sum of the directional flow rates across sections perpendicular to the main flow), we perform a comparison at fixed density. Our data (inset of Fig.~\ref{fig:risk_vs_density}) reveal only little benefit to switching from two-way to one-way traffic in our wide-path scenarios. 
To further test this somewhat surprising finding, we exploit the \emph{controlled} experiments performed a few years ago by the German BaSiGo team\footnote{These extensive datasets are openly available under: https://ped.fz-juelich.de/db/doku.php} to study unidirectional and bidirectional pedestrian flows in 4 to 5-meter-wide corridors \cite{cao2017fundamental}; head orientations are reconstructed as explained above. The results, shown in the top left inset of Fig.~\ref{fig:risk_vs_density}, are in line with the aforementioned finding: In wide walkways, switching from two-way flow to one-way flow seems to 
(at best) reduce the risks only moderately, probably because head-on 'collisions' are rare in these self-organized flows.

Next, we turn to queues and study how their arrangement affects transmission
risks. On the basis of our observations at a testing site, 
we modeled a queue as a line of more or less
equally spaced people, swaying 
in a $50\,\mathrm{cm}\times50\,\mathrm{cm}$
rectangle around their central spot
and whose head orientations are
normally distributed (with a standard
deviation of $22^{\circ}$) around the queuing axis 75\% of the time and purely random for
the remaining 25\% of the time, due to people turning around or having
a look around. Both the positions and
orientations are refreshed every second.
Albeit simplistic 
\footnote{The observed scenario is significantly more complex than its 
reconstruction: It actually features two different,
not strictly linear queues, one outdoors and one indoors, as well as a few people who are not queuing.},
the reconstructed queuing scenario is comparable with our actual observations as far as the estimated
infection rates are concerned; these estimates with and without reconstruction differ by less than 50\% for any of our models -- except the most optimistic ones, which bestow special importance
to rare contact events that are overlooked in the reconstitution.

\begin{figure}
\begin{centering}
\includegraphics[width=0.5\textwidth]{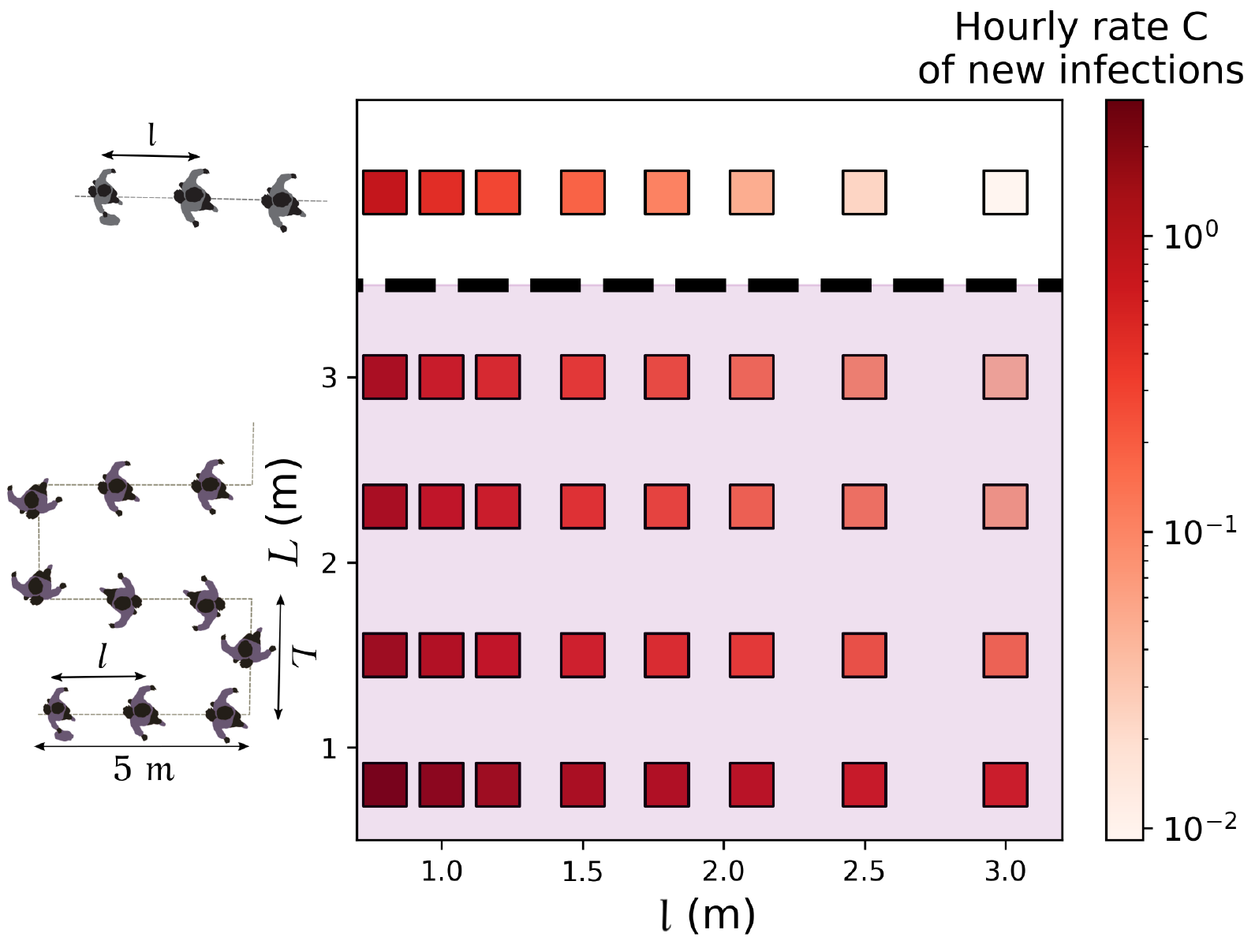}
\par\end{centering}
\caption{\label{fig:queue}Hourly rate of new infections in a linear queue
(top row) and a winding queue, depending on the spacing between pedestrians
and lines, as estimated with the \emph{ModOpt}$_{3}$ parameter set. People
move one spot forward every other minute. Note the logarithmic color
scale on the graph.}
\end{figure}

Figures~\ref{fig:queue} and S7 illustrate the extent to which predicted infection rates
vary when the spacing between queuing people or the
queuing geometry are modified. 
\revise{}
{
Naturally, the risks are minimal in the case of the linear queue with the largest spacing between individuals, whereas they are maximized for winding, S-shaped queues with very close rows and
individuals standing right behind each other within each row. Since people cannot be expected to keep their heads strictly in the direction of the queue, there appears to be no
smart solution to have at the same time minimal risks and a very compact queue; given the results of Fig.~\ref{fig:queue} and S7, a simple practical recommendation for S-shaped queues is to keep a little more
space between rows than the actual spacing between individuals in each row.
}

\subsection{Current limitations and perspectives}
\hfill \\
\revise{}{
In summary, the foregoing risk assessments in non-confined environments can guide
public decisions in times of pandemic, in that (irrespective of the
transmission parameters that are used) they confirm the risks of infection
incurred at caf\'es \cite{chang2020mobility} and underline the key
role of pedestrian density in determining the rate of viral transmission
in moving crowds without masks. Fairly busy streets, with densities up to $\rho\simeq0.1\,\mathrm{ped/m^{2}}$,
are found to present risks that are not completely negligible, but comparatively quite
low, and these risks will manifestly reach even lower values for less
busy streets. This suggests that the scant reports of
outbreaks in outdoor walking crowds are not only due to the intricacy of
tracing back these infections (due to unidentified contacts, recall
biases, etc.), but also to the limited transmission of the virus in
these conditions, even without face coverings. Nevertheless, this
remark does not apply to very crowded settings
such as markets or
metro and train stations, which deserve particular attention.
Furthermore, our model-based approach has enabled us to explore the  efficiency of street and venue redesigns in mitigating the viral spread.
For wide walkways, we have not found clear benefits to switching from two-way foot traffic to one-way traffic. For queues, increasing the space between
individuals naturally reduces transmission risks; in the case of an S-shaped queue, a simple rule that could be enforced in practice is to keep a little more space between
rows than between individuals in each row. 
}

Given that our study pioneers the coupling of empirical crowd data to spatial models
of viral transmission at mesoscales, it undoubtedly suffers
from some limitations. To start with, the empirical data could be
extended to include more scenarios and longer footage.

Perhaps more crucially, the transmission model should be
refined. The models used in this study are admittedly overly simple,
even though this problem was partly warded off by ensuring the robustness
of our qualitative conclusions across diverse model variants, including a spatio-temporal model. More sophisticated
models, which may differentiate transmission rates as a function of
people's activity (reflecting known variations in droplet emission \cite{asadi2019aerosol,abkarian2020speech})
and account for the effect of the wind \cite{feng2020influence} and ambient air flows, will afford more accurate
estimation of the rate of new infections. In addition, fluid dynamics
simulations of long-range aerosol propagation would make it possible
to study enclosed spaces with poor ventilation, where our current
models that discard the airborne transmission route can only provide
lower bounds on infection risks.

Another task is to generalize the transmission
models to people who are (adequately or inadequately) wearing a mask
\cite{leung2020respiratory}, in order to determine how serious an
issue very crowded streets really are in current times. It would be
straightforward to account for the particle filtration efficiency of masks
in the present framework, by simply multiplying the transmission rates
by a reduction factor (say, $\sim20\%$ for cloth masks, $\sim10$\%
for surgical masks and $\ll5\%$ for N95 masks \cite{li2020assessing,bar2020quantitative},
if \emph{only the emitters} have their face covered), but 
masks are probably even more efficient, because they also reduce the reach of
the exhaled puff \cite{bahl2020face,bhagat2020effects}, thereby probably shortening
the range of transmission of droplets \cite{li2020assessing}.

 \section*{Acknowledgments}
 We are grateful to Isabelle Sabran (Ville de Lyon), Jessica
 Magraner (Cour d'appel de Lyon), and Fr\'ed\'eric Laurent (Hospices Civils
 de Lyon) for facilitating our collection of data with our privacy-respective
 cameras and to the French MODCOV19 initiative for supporting
 part of this work. We also thank C\'ecile Appert-Rolland for lending us some material and Marina Nicolas for proofreading.
 This work was funded by Agence Nationale de la Recherche (ANR-20-COV1-0003) under project name \emph{SeparationsPietons}. 
 The setup of the CFD simulations discussed in Appendix D was designed collectively, with P. B\'enard, G. Lartigue, V. Moureau (CORIA Rouen, France), G. Balarac, P. B\'egou (LEGI Grenoble, France), Y. Dubief (Univ. Vermont, USA) and R. Mercier (Safran Tech, France). CFD simulations were performed using HPC resources from GENCI-TGCC. SM also acknowledges the support of Agence Nationale de la Recherche (ANR-21-CO15-0002, TransporTable).

\newpage

\setcounter{equation}{0}
\renewcommand{\theequation}{S\arabic{equation}}
\setcounter{table}{0}
\renewcommand{\thetable}{S\arabic{table}}
\setcounter{figure}{0}
\renewcommand{\thefigure}{S\arabic{figure}}%

\appendix
\onecolumn

\section{\label{ref:SI_empirical_observations}Empirical observations of pedestrian crowds}

\subsection*{Video acquisition and processing}

Videos of crowds were acquired from a top view in the non-confined settings described in Table~1, in a discrete and passive
way that allowed us to observe natural
behavior as pedestrians were mostly unaware of being filmed.

{\small
 \begin{centering}
\begin{longtable}[h]{|>{\centering}m{3.5cm}|>{\centering}m{3.2cm}|>{\centering}m{3.8cm}|>{\centering}m{2.3cm}|>{\centering}p{1.8cm}|}

\hline 
Scenario / Location & Date and time & Comments & Number of tracked pedestrians & \textbf{Mean Density (ped/m$^{2}$)}\tabularnewline
\hline 
\hline 
Pedestrian banks of the Rhône river, close to the Morand Bridge & Friday, July 3rd 2020, 4pm-5pm & Sunny and windy day & 164 & \textbf{0.042}\tabularnewline
\hline 
\textcolor{black}{Plaza in front of Perrache hub
(Hall - Level 1)} & Friday, July 17th 2020, 9am-10am & Nice weather & 1021 & \textbf{0.038}\tabularnewline
\hline 
Part-Dieu train station - Ground level/passage area (indoors) & Friday, July 17th 2020, 11:20am-12:40pm & Sunny day & 875 & \textbf{0.22}\tabularnewline
\hline 
Busy street - Under the Passerelle Bouchut & Wednesday, July 29th 2020, 12pm-1pm & Hot and sunny day & 800 & \textbf{0.05}\tabularnewline
\hline 
Bellecour subway station - Platform of Line D (indoors) & Tuesday, July 28th 2020, 3pm-4pm & Hot and sunny day & 849 & \textbf{0.26}\tabularnewline
\hline 
Croix-Rousse boulevard - street caf\'es & Friday, September 11th 2020, 6:30pm-7pm & Nice weather & 13 & \textbf{/}\tabularnewline
\hline 
Grande rue de la Croix Rousse (shopping street) & Saturday, January 16th 2021, 11am-12pm &  Cold and cloudy day & 420 & \textbf{0.06}\tabularnewline
\hline 
Saint-Jean street in the Old Town of Lyon & Monday, September 21st 2020, 1pm-2pm & Hot and cloudy day & 481 & \textbf{0.11}\tabularnewline
\hline 
Place des Terreaux - Bar/Restaurant terraces & Thursday, October 8th 2020, 5:40pm-6pm & Sunny day, cool weather & 30 & \textbf{/}\tabularnewline
\hline 
Croix-Rousse - Main market alley & Sunday, October 18th 2020, morning & Sunny day & 183 & \textbf{0.46}\tabularnewline
\hline 
COVID-19 testing site & Monday, October 19th 2020, 8:30am-10:30am & Outdoor waiting lines (cold day) - Indoor waiting lines (sport arena) & 66 & \textbf{/}\tabularnewline
\hline 

\par
\caption{\label{tab:SI_scenarios}Details about the investigated scenarios. All sites are in the metropolitan area of Lyon, France; most are outdoors.}
\end{longtable}
\end{centering}
}

After correcting for lens deformation, the trajectories and head orientations were extracted by extensive manual tracking 
with the help of a dedicated Python script and a touch screen. The pixel coordinates $(x^{\prime},y^{\prime})$ are converted into real-world coordinates $(X^{\prime},Y^{\prime})$ using the following geometric relation,

\begin{eqnarray*}
X & =X_{c}+\gamma\left(\nu x-X_{c}\right)\\
Y & =Y_{c}+\gamma\left(\nu y-Y_{c}\right)
\end{eqnarray*}
where $\gamma=\frac{1}{1-\sin\alpha\cdot\frac{y}{D}}$, $\alpha$ is the angle between the camera axis and the vertical direction (we systematically found $\alpha\simeq 0$, reflecting the top view used in virtually all videos). Here, 
the $(X,Y)$ and $(x,y)$ coordinates are defined in a frame that has been conventiently rotated in the horizontal plane so that one of its axes coincides with the horizontal projection of the camera axis; these coordinate systems are connected to $(X^{\prime},Y^{\prime})$ and $(x^{\prime},y^{\prime})$, respectively,
by rotations in Euclidean space.
The unknown parameters in this relation (i.e., the angles of these rotations, $\nu$, $\gamma$, $X_c$ and $Y_c$) are estimated by fitting at least 4 calibration points at predefined positions, where
a team member stood in each scenario.

The expected uncertainty on positions created by failing to account
for height differences ($\delta h$) between individuals is expected
to be around $\delta x\simeq\tan\alpha^\prime\cdot\delta h\approx15\,\mathrm{cm}$
in the present conditions, where $\alpha^\prime$ is the angle in which 
the pedestrian was
filmed with respect to the vertical. In practice, by double-tracking some test
pedestrians, we estimate that the uncertainty is typically below or
around 20 cm. The same method allowed us to evaluate the standard
deviation of the error on head orientations to $19^{\circ}$. Some
videos, especially for the static scenarios, were filmed from a more
distant viewpoint, in which case the error on the orientations is
likely to be larger.

\begin{figure}
\begin{centering}
\includegraphics[width=0.45\textwidth]{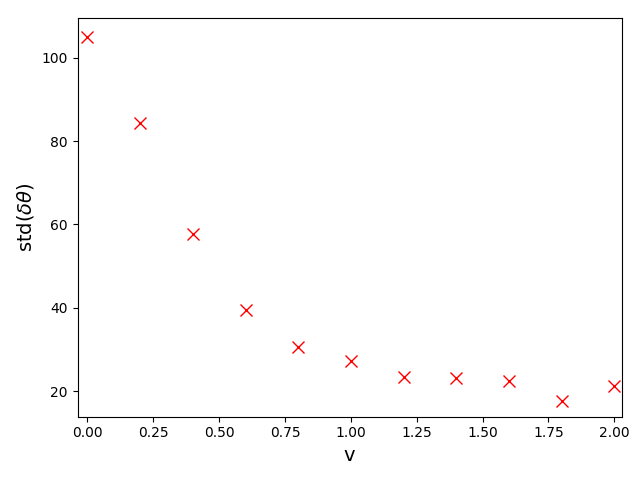}
\par\end{centering}
\caption{\label{fig:deltaTheta_PD} Standard deviation of the angle $\delta\theta$ between the walking direction and the head orientation as a function of speed, for the data collected at the \emph{Part Dieu} railway station.}
\end{figure}

The accuracy of the head orientations extracted from the videos, especially in moving crowds, is also supported by the study of the angular difference $\delta\theta$
between the head orientations and the walking direction of pedestrians. Since we expect that $\langle \delta\theta \rangle=0$ on average, we are interested in the standard deviation of $\delta\theta$, which we plot
as a function of the walking speed in Fig.~\ref{fig:deltaTheta_PD}. We observe that 
head orientations tend to align more and more with the walking direction of pedestrians as they walk faster and the standard deviation smoothly reaches values around $20^{\circ}$ at relatively high speeds, which confirms the accuracy of our tracking.

\subsection*{\label{sec:SI_rescaling}Contact rescaling to compensate for undetected interactions}

The finite field of view of the camera results in some missed interactions with off-camera people.
We can estimate the fraction $\phi(r)$ of missed contacts depending on the separation distance $r$ under an assumption
of uniform density. This is achieved numerically by randomly placing points (pedestrians) in an area with the same dimensions as the view field and testing if secondary points inserted at a distance $r$ of the first one in a random direction are inside or outside the area. Then, all detected interactions are multiplied by a factor $\frac{1}{1-\phi(r)}$ to compensate for those that went undetected; for practical purposes, we set a maximal value (20) to this factor, to avoid excessive
amplification of the statistical noise at large separation distances $r$.

\begin{figure}
\begin{centering}
\includegraphics[width=0.4\textwidth]{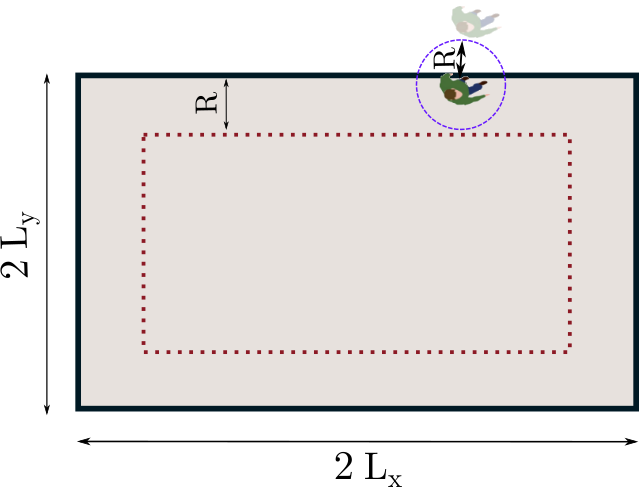}\ \ \
\includegraphics[width=0.45\textwidth]{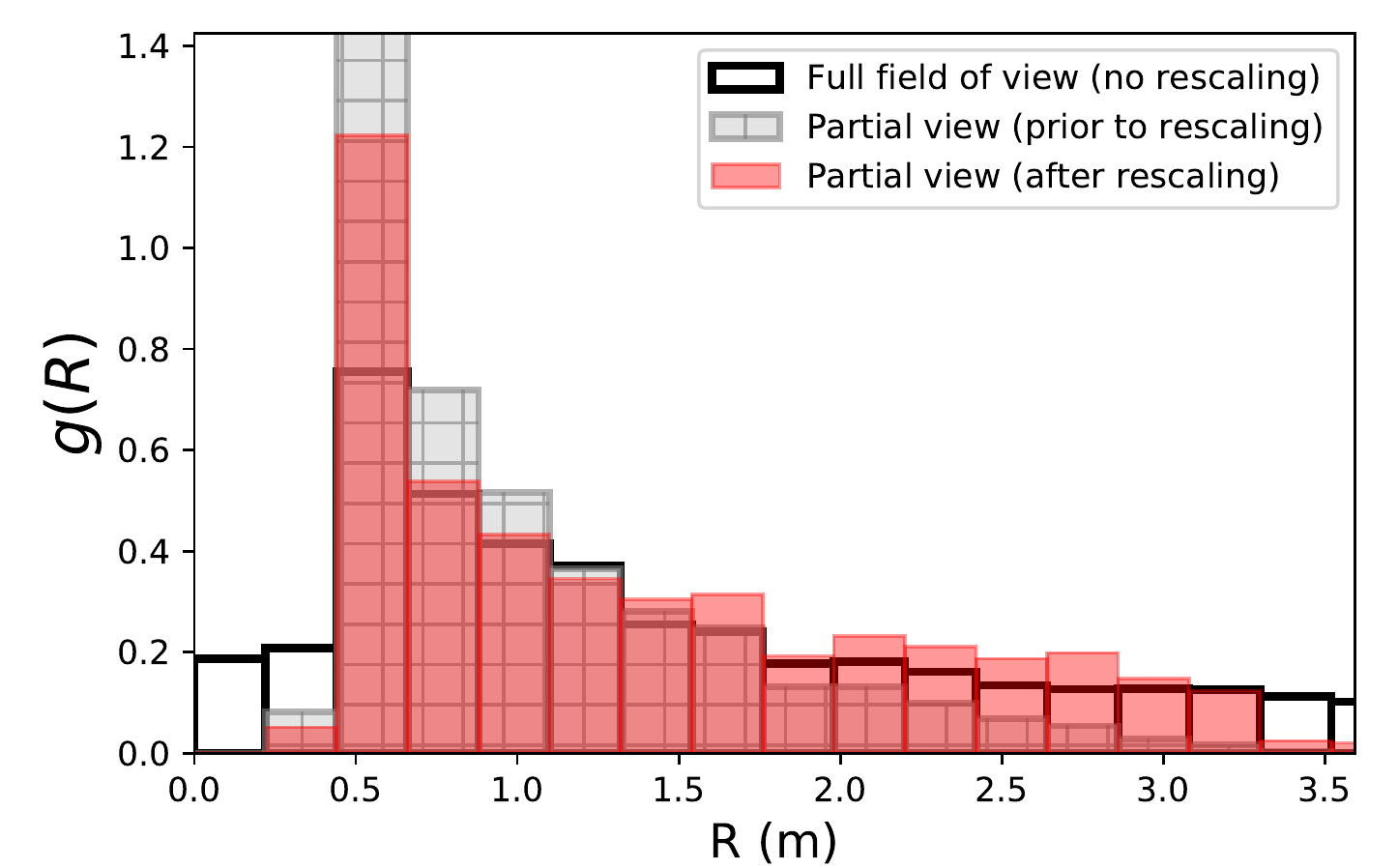}
\par\end{centering}
\caption{\label{fig:SI_RDF}Contact rescaling method. (\emph{Left}) Sketch illustrating the possibility to miss interactions with off-camera people. (\emph{Right}) Radial distribution function $g(r)$ in the Perrache plaza scenario measured using the original
field of view or a reduced field with and without application of the contact rescaling method.}
\end{figure}

This method is put to the test by focusing on a small area of dimensions 3~m$\times$3~m in the view field of a given scenario
and computing the radial distribution function $g(r)$ of contacts (i.e., $g(r)\propto \frac{N(r)}{2\pi r}$ where $N(r)$ is the number of contacts of range $r$) in the 'partial' field before and after applying the rescaling method. Figure~\ref{fig:SI_RDF} demonstrates the efficiency of the method: After rescaling, $g(r)$ gets much closer to the distribution measured in the original field of view (up to statistical noise, since only a fraction of the people are then observed).

\section{\label{sec:SI_transmission_models}\emph{Ad hoc} models for viral transmission}

This supplementary section details and discusses the \emph{ad hoc}
functions that were introduced to model the transmission rate $\nu$,
defined such that $\nu dt$ is the probability for a \emph{susceptible}
individual to be infected by the index patient over the interval $dt$.

\subsection*{Specification of transmission models}

We start by recalling the generic expression chosen for these functions
(Eq.~3 of the main paper),

\begin{equation}
\nu(r,\theta^{E},\theta^{R})=\frac{1}{\tilde{T}_{0}}\bar{f}\left(\frac{r}{r_{0}}\right)\cdot\bar{f}\left(\frac{\theta^{E}}{\theta_{0}^{E}}\right)\cdot\bar{f}\left(\frac{\theta^{R}}{\theta_{0}^{R}}\right),\label{eq:SI_nu_expression}
\end{equation}
emphasizing that $\theta^{E}$ and $\theta^{R}$ are the orientations
of the pedestrians' heads, hence mouths and noses, relative to the
vector that connects them, in the \emph{horizontal plane}, i.e., irrespective
of the difference in height. Accordingly, $\theta^{E}=\theta^{R}=0$
corresponds to a face-to-face interaction.

The formulation of Eq.~\ref{eq:SI_nu_expression} is 
arbitrary to a large extent, but its main features are derived from experimental
and numerical data. Experimental images of inhalation \cite{laverge2014experimental}
and expiratory emissions during breathing, speaking \cite{abkarian2020speech},
or coughing \cite{bourouiba2014violent}, as well as computational
fluid dynamics simulations \cite{chen2020short,feng2020influence,abkarian2020speech},
brought to light strong \emph{directed} transport of respiratory droplets
overs tens of centimeters, reaching a couple of meters; the exhaled
cloud is \emph{mostly enclosed} in a cone of half-angle between $10^{\circ}$
and $50^{\circ}$ at most, depending on the speech conditions \cite{abkarian2020speech},
thus pointing to a significantly smaller characteristic angle $\theta_{0}^{E}$.
In passing, note that such conical propagation is well described by Eq.~\ref{eq:SI_nu_expression},
in which $r$ and $\theta^{E}$ are decoupled. Inhalation is also
directional, but, as ambient air from all around the mouth will be
breathed in \cite{abkarian2020speech}, we have systematically considered that its
directionality was less marked, i.e., $\theta_{0}^{R}>\theta_{0}^{E}$ (see Table~1 of the main paper).

Regarding the characteristic distance $r_{0}$, a review of 172 observational
studies \cite{chu2020physical} tentatively suggests that keeping
a physical distance of 1 meter reduces the infection risk from 13\%
(very close contact) to 3\% and down to 1.5\% if a distance of 2 meters
is maintained, although there is very large dispersion in the data.
In the UK, the Scientific Advisory Group on Emergencies (SAGE) asserts
that physically distancing by 2 meters instead of 1 reduces the risks
by a factor between 2 and 10 \cite{SAGE_Britain}. This translates
into an exponent $m\approx2-3$ in function $\bar{f}_{2}$, a value
$r_{0}\in[0.4\,\mathrm{m},1.4\,\mathrm{m]}$ if one uses function
$\bar{f}_{3}$ or $r_{0}\in[1\,\mathrm{m},2\,\mathrm{m]}$ if one
uses function $\bar{f}_{1}$. However, at larger physical distances,
contacts may have been less in a face-to-face orientation or shorter
in duration, so these ranges ought to be taken with a grain of salt.
Chen et al.'s theoretical work rather suggests $r_{0}\approx0.2\,\mathrm{m}$
for talking and $\mathrm{r_{0}}\approx0.5\,\mathrm{m}$ for coughing
(see Fig.~7 of \cite{chen2020short}). 
In any event, our goal is to consider a family of parameter sets spanning the whole range between those for which viral transmission is more likely (conservative, or `pessimistic', approach) and those for which it more seldom occurs (`optimistic', i.e., predicting fewer infections). 

As for the time parameter, it is most easily expressed as $T_{0}=\tilde{T}_{0}/\bar{f}(\frac{0.5}{r_{0}})$
which is the characteristic duration for infection at a distance $r=0.5\,\mathrm{m}$.
Contact tracing routines generally consider a threshold duration between 5
and 15 minutes. Nonetheless, this value may be at the safe end of
the spectrum, as exposure case studies have only reported very few
cases with contacts lasting less than 15 minutes.

Finally, we also modeled uncovered sneezes, which are longer ranged
than other expiratory events but also highly directional \cite{bourouiba2014violent}.
Even though the emitted jet is reported to reach 7 or 8 meters, not
all droplets are expected to travel this far, as seen in the experimental
images of \cite{bourouiba2014violent}. Accordingly, we settled for
$r_{0}=1.5\,\mathrm{m}$, which means that the risk at a distance of 7 meters
is 1\% of that at 50 cm with the $\bar{f}_{3}$ function, 3\% with $\bar{f}_{2}$,
but close to zero with $\bar{f}_{1}$.

There is no denying that the foregoing \emph{ad hoc} models and their
parameters are but coarse approximations of the reality: They do not
account for the effect of wind, ventilation, or humidity \cite{feng2020influence},
nor do they describe the dynamics of exhaled jets and the time lag
due to their propagation. Perhaps more importantly, the variability
in droplet emission and viral shedding between individuals is overlooked,
as is their dependence on their activity (breathing through the nose
or the mouth, panting, talking loud, coughing, etc.), which is known
to have a major impact \cite{fennelly2020particle,morawska2009size,asadi2019aerosol}.
However, since virtually all scenarios under consideration (with the
exception of the caf\'es) involve similar activities, i.e., very moderate
physical activity and limited talking, we expect our approach to hold
in a \emph{statistical} sense, although different index patients may
be associated with different parameter sets in the model. The characteristic
duration $T_{0}$ is set by noting that a single sneeze may expel 10 times as many
respiratory droplets as 5 minutes of talking \cite{poon2020soft},
but generally does not occur more than once every few minutes. This
is particularly true if the focus is on \emph{uncovered} sneezes.
Accordingly, as a \emph{worst-case} estimate, we set $T_{0}=100\,\mathrm{s}$.

\subsection*{Consistency with the current knowledge about infection risks and
droplet emission}

Let us now examine to what extent case reports may support the proposed
models.

Generally, outbreak reports do not specify the very precise
circumstances of the infections. Still, the attack rates in diverse
settings are useful indications. Indeed, within households, reported attack rates range from 5\% to
30\% \cite{bar2020quantitative}, for instance around 15\% in \cite{park2020early,park2020contact,burke2020active}. At work or in the community, 
casual episodic
contacts, even face to face, do not necessarily trigger an outbreak
of cases \cite{park2020early,burke2020enhanced}. Furthermore, it
was reported that at Solano County hospital (California, US) the majority
of healthcare personnel did not get infected despite spending 10 to
50 minutes in the same room as a Covid patient, often within 2 meters
and with no facial mask \cite{heinzerling2020transmission}. All
these pieces of evidence hint at an average time for infection $T_{0}$
in a close face-to-face contact that is probably longer than $\sim10$
minutes.

Another general indication comes from the basic reproductive number
$R_{0}$ in the pre-pandemic context, which was around $R_{0}\approx3$.
Contact pattern data in this usual, pre-pandemic context imply that
only 7\% of contacts lasting longer than 15 minutes lead to an infection \cite{moritz2020risk}.
To translate this probability of infection into parameter values for our model, we need to
make some assumptions about how long and how close these interactions were actually.
We shall assume that prolonged contacts were typically 20 minutes long in a face-to-face setting and we
will leave aside possible differences in height.
Then, if the contact distance is only 50 cm, this yields a characteristic time for infection $T_{0}$ of a couple of hours.
If the contact distance is only 80 cm, $T_{0}$ is found to be between 20 minutes
and 2 hours, depending on the functional
choice $\bar{f}_{1}$, $\bar{f}_{2}$ or $\bar{f}_{3}$ and the chosen
value for $r_{0}$. Thus, this gives credence to the `optimistic'
end of our parameter spectrum.

\begin{figure}
\begin{centering}
\includegraphics[width=7cm]{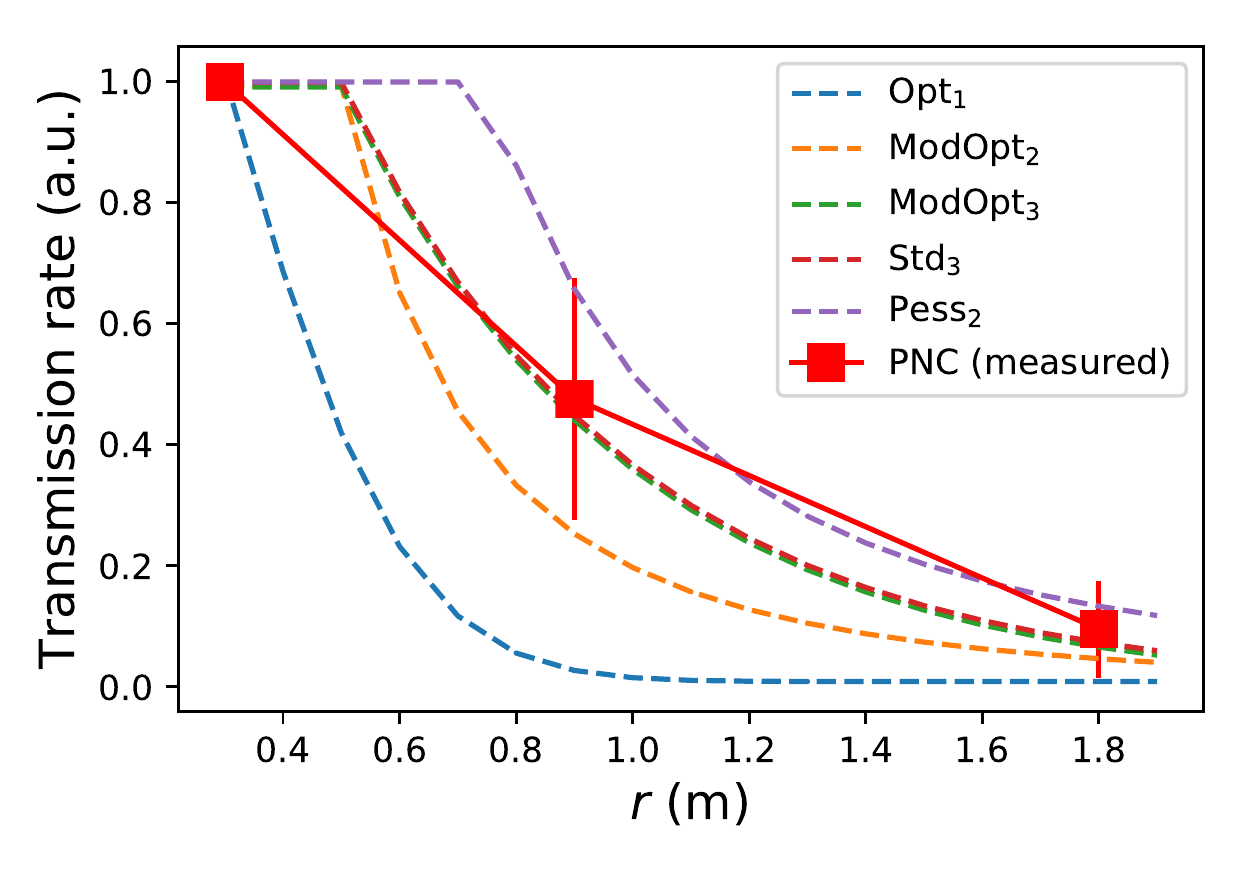}
\par\end{centering}
\caption{\label{fig:SI_PNC_decay}Comparison between the spatial decay of the
modeled transmission rate $\nu(r,\theta^{E}=0,\theta^{R}=0)$ with
distance $r$ and the particle number concentration (PNC) measurements
during coughing of \cite{li2020assessing}. Arbitrary units have
been chosen so that the value at $r=0.\,\mathrm{m}$ is always unity.\protect \\
The following abbreviations are used: \emph{Opt$_{1}$ }(optimistic
parameters combined with $f_{1}$), \emph{ModOpt$_{2}$ }(moderately
optimistic parameters combined with $f_{2}$), \emph{ModOpt$_{3}$
}(moderately optimistic parameters combined with $f_{3}$), \emph{Std$_{3}$
}(standard parameters combined with $f_{3}$), \emph{Pess$_{2}$ }(pessimistic
parameters combined with $f_{2}$).}
\end{figure}

Further insight is provided by empirical measurements of the particle
number concentration of the respiratory droplets exhaled by a (healthy)
subject \cite{li2020assessing}. The concentration of droplets of
sizes $0.5\,\mathrm{\mu m}-20\,\mathrm{\mu m}$ was measured at different
distances in front of a subject during coughing, with and without
face covering. In Fig.~\ref{fig:SI_PNC_decay}, we compare the spatial
decay of these measurements (without face covering) to the predictions
of some of our transmission models. Excellent agreement is found with
the moderately optimistic and standard models combined with $f_{3}$.
Yet, one should bear in mind that the measurements were made while
the subject was coughing, which may increase the transmission range.
Therefore, these data once again support the most optimistic of our
transmission models.

\subsection*{\label{sec:SI_order_magnitude}Order of magnitude of droplet emission}

One may also wonder how compatible the model parameters
are with the current fundamental knowledge about droplet emission. Exhaled droplets
are very broadly distributed in size around a mean value of $\sim5\,\mathrm{\mu m}$
and contain $n_v \sim 0.1$ viral copy (mostly zero or one copy) per droplet
\cite{poon2020soft}. Talking for five minutes, or coughing once
in these 5 minutes (which is a relevant frequency according to coughing
statistics \cite{bourouiba2020fluid}) will expel about $N_d\approx 3\cdot10^{3}$
droplets. It is difficult to estimate what fraction of these will be breathed
in by a person standing 50 cm apart, but, following the reasoning
of \cite{abkarian2020speech}, droplets should be diluted by a factor
$\eta_{\mathrm{dil}}\sim0.1$ at this distance, so it is reasonable to estimate that
at most a few $\%$ ($\eta_{\mathrm{inh}}$) of the droplets will be inhaled, a significant
fraction ($\eta_{\mathrm{naso-ph}}$) of which may deposit in the nasopharynx, where they are most
likely to lead to an infection \cite{basu2020close}. Assuming an
infectious dose of $n_\mathrm{inf} = \mathcal{O}(100)$ particles \cite{basu2020close}, i.e.,
that only one in a hundred virions will successfully invade a host
cell and replicate, one arrives at a characteristic time for infection
$T_{0}=\frac{n_\mathrm{inf} \cdot 5\,\mathrm{min}}{N_d \cdot n_v \cdot \eta_{\mathrm{dil}} \cdot \eta_{\mathrm{inh}} \cdot \eta_{\mathrm{naso-ph}}}$ of several hours $(T_{0}\lesssim 10\mathrm{h}$). Notwithstanding
how rough an estimate this is, it is worth noticing that it is once again
rather in line with the very optimistic end of our spectrum of plausible
values.

As a matter of fact, on the basis of a similar, but more rigorous reasoning, Yang et al. derived
shorter infection times $T_0$, but their explicitly conservative approach assumes constant speech
and rests on the idea that any inhaled droplet reaches zones where viral penetration in the cellular tissue is possible \cite{yang2020towards}.

\subsection*{\label{sec:SI_Chinese_train}Direct comparison with the collated statistics of infections aboard Chinese trains}

It is generally difficult to test transmission models against empirical
data related to Covid-19 cases in a statistically meaningful way because
the detailed interactions between the index patient and his or her
contacts are seldom known. However, Hu et al.'s study of transmission
risks among train passengers in China \cite{hu2020risk} provides
a virtually unique opportunity to attempt such a direct comparison.
Indeed, Hu et al. were able to retrieve the trip record of confirmed
Covid-19 patients who had traveled on a train in the 14 days before
the onset of the illness, between December 2019 and late February
2020. Having access to the detailed seating plan on their train coaches,
the researchers then tracked reported Covid-19 cases among co-travelers
and computed the infection probability depending on the distance (expressed
as a number of rows and columns) between the index patients and the
contacts. A marked increase in the infection risk was found for people
seated in the same row as the index patient, especially in adjacent
seats, for more than a couple of hours.

\begin{figure}
\begin{centering}
\begin{subfigure}
Optimistic parameters combined with $f_{2}$:
\includegraphics[width=0.9\linewidth]{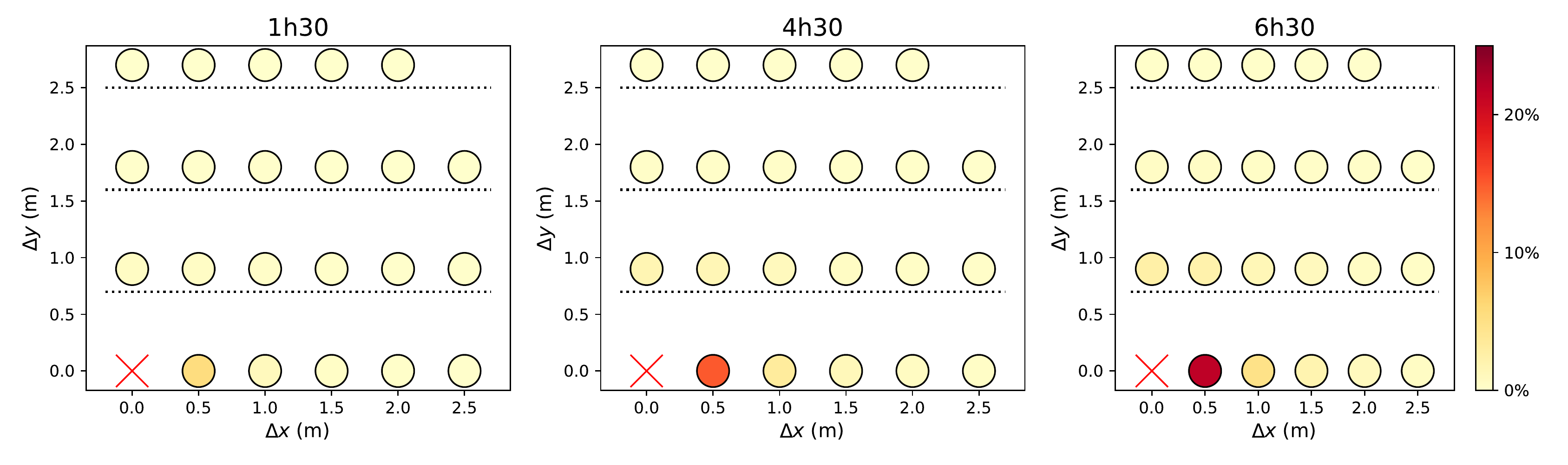} 
\end{subfigure}
\begin{subfigure}
Moderately optimistic parameters combined with $f_{1}$:
\includegraphics[width=0.9\linewidth]{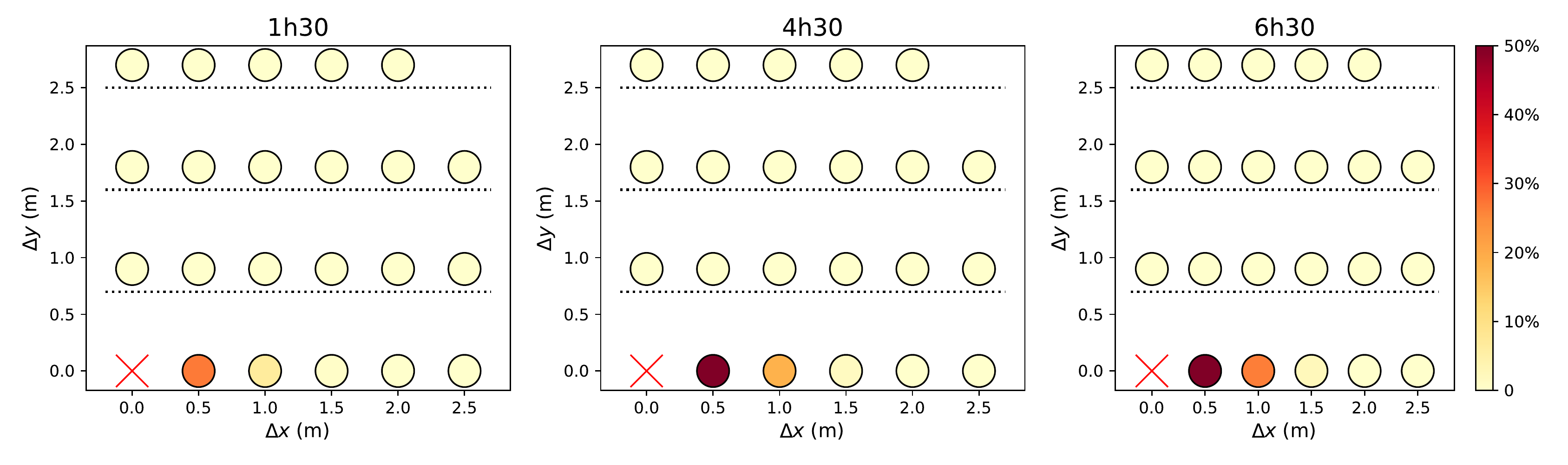}
\end{subfigure}
\begin{subfigure}
Moderately optimistic parameters combined with $f_{3}$ (\emph{ModOpt}$_{3}$):
\includegraphics[width=0.9\linewidth]{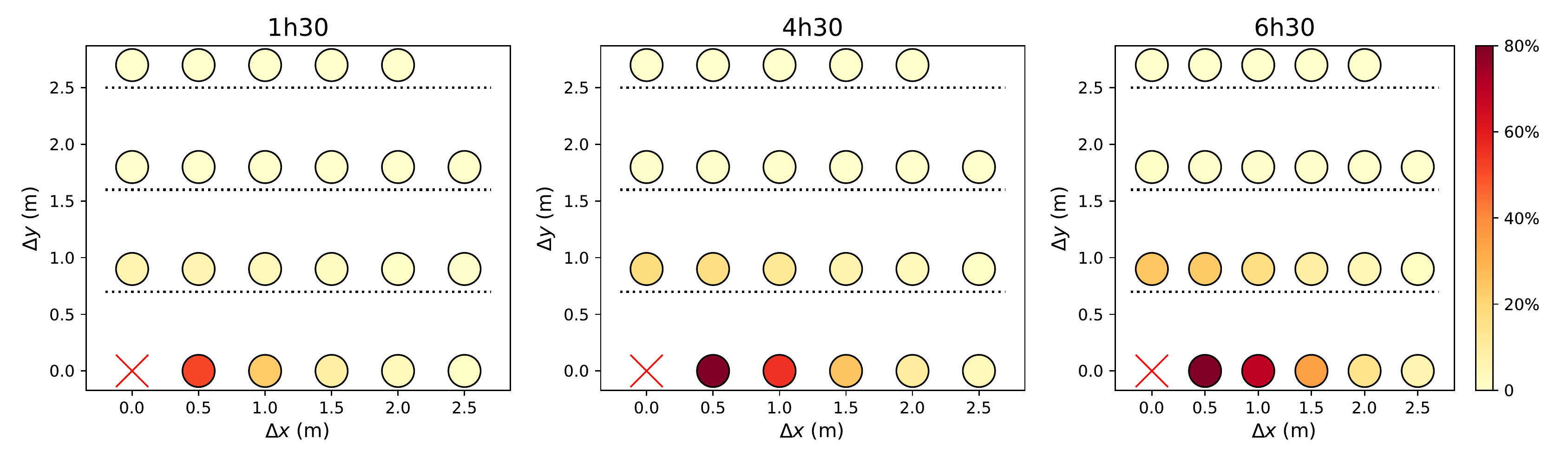}
\end{subfigure}
\par\end{centering}
\caption{\label{SI_fig:train_infections}Simulated probability of infection
of co-passengers aboard the train, depending on their proximity to
the index patient (\emph{red cross}) and the time spent together on
the train. $\Delta x$ denotes the \emph{absolute} distance along
each row and $\Delta y$ denotes the distance along each column; the
physical barriers created by seats are materialized as a dashed line.
Model parameters are specified above each row of charts. These figures are amenable to a direct comparison with Fig.~2 of \cite{hu2020risk}.}
\end{figure}

Leaving aside the fact that a train coach is an enclosed (but usually
ventilated) space, the transmission models introduced above can be
applied to these settings. Distances between seats
and rows are precisely known. The central aisle is counted as
one seat, as in the original study \cite{hu2020risk}, while the
barrier to transmission created by seat rows (and represented by dashed
lines in Fig.~\ref{SI_fig:train_infections}) is arbitrarily considered
to have an effect comparable to an additional distance of 1 meter.
In the light of our empirical measurements
on moving and queuing crowds (see the main text), the passengers'
head orientation is assumed to be normally distributed around the
axial direction, with a standard deviation of $26^{\circ}$; we consider that the seated
position suppresses the rotations of the whole torso
that are occasionally observed in standing pedestrians.

Figure~\ref{SI_fig:train_infections} presents the probabilities
of infection evaluated on this basis using three different transmission
models on the optimistic side of the parameter spectrum, with the
help of Eq.~\ref{eq:evolution_Ij_2}. These results are directly
confronted with Fig.~2 of \cite{hu2020risk}; the three of them are found to
compare well as far as both the evolution of the risks with co-travel
duration and their spatial pattern are concerned.

A note of caution
should nevertheless be made about the absolute values of infection probabilities,
which peak at about 10\% in \cite{hu2020risk}, compared to 25\%
to 80\% here. This very notable difference may largely be due to
the fact that all trips of the index patient in the two weeks before
his or her diagnosis are taken into account in \cite{hu2020risk} whereas
it is very unlikely that the patient was contagious during all this
period. If the contagious period before the onset of symptoms spans
two to three days \cite{cevik2020sars,bar2020quantitative}, i.e.,
about 15\% to 20\% of the two-week period, then applying this 15-20\%
ratio to the simulated infection probabilities yields values that
become comparable to those reported by \cite{hu2020risk}. These
pieces of evidence thus give credence to our transmission models coupled
with optimistic parameter sets.

Another caveat should however be mentioned regarding Hu et al.'s work \cite{hu2020risk}:
The researchers admit that they were not able to distinguish relatives from unrelated
people in their data. Therefore, part of the reported infections may
not have taken place aboard the train, but elsewhere.

\clearpage

\section{\label{sec:SI_global_rate}Transmission}

Our study relies on the independent action hypothesis introduced by
Druett in the early 1950s \cite{druett1952bacterial} and partially
validated by Zwart in virus-insect pathosystems a decade ago \cite{zwart2009experimental}.
It posits that there is no minimal infectious dose and that one can
overlook cooperative or antagonistic interactions between virions
in the host system and assign to each of them a finite probability
(here denoted by $\epsilon$) to cause an infection. This yields a Wells-Riley-like equation \cite{sze2010review} for the time evolution of the contagion status $I_{j}(t)$
of (initially susceptible) agent $j$, Juliette, infected by index patient $i$, Iago ($I_{j}$ is the probability that
Juliette is infected), viz.,

\begin{eqnarray}
\dot{I}_{j}(t) & = & \left[1-I_{j}(t)\right]\nu_{ij}(t),\label{eq:evolution_Ij-1}\\
I_{j}(t) & = & 1-e^{-N_{ij}(-\infty,t)},\,\mathrm{with}\,N_{ij}(s,t)=\int_{s}^{t}\nu_{ij}(t^{\prime})\,dt^{\prime}\label{eq:evolution_Ij_2}
\end{eqnarray}
where $\nu_{ij}(t)=n_{ij}(t)/n_\mathrm{inf}$ is the risk transmission
rate, $n_\mathrm{inf}$ is the minimal infectious dose, and $n_{ij}(t)$ is the number of viral copies exchanged between Iago and Juliette per unit time around time $t$. Accordingly,
if Iago was filmed between time $t_{0}$ and time
$t_{0}+\tau_{i}$, the number of new infections that were actually witnessed (or 'event $R$' in the terminology of \cite{tupper2020event})
can be estimated to
\begin{eqnarray}
C_{i}^{(\tau_{i})} & = & \sum_{j\notin G_{i}}I_{j}(t_{0}+\tau_{i})-I_{j}(t_{0})\nonumber \\
 & = & \sum_{j\notin G_{i}}S_{j}^{0}\cdot\left[1-e^{-N_{ij}(t_{0},t_{0}+\tau_{i})}\right]\label{eq:C_i_with_S0}
\end{eqnarray}
where $G_{i}$ contains Iago and all other agents that we visually
identified as members of his group (family, friends, etc.) and $S_{j}^{0}=1-I_{j}(t_{0})=e^{-N_{ij}(-\infty,t_{0})}$
is the probability that Juliette was already
infected at $t_{0}$. Note that similar assumptions about the rate of infections were made in a recent paper that we 
came across just before submission \cite{tupper2020event}. Let us emphasize in particular that intra-group
infections are discarded here, because group members are likely to
have been infected outside the scenario of interest, whereas our purpose is
to estimate \emph{new} infections in the scenario.

\subsection*{Scenarios involving a moving crowd}

To compute Eq.~\ref{eq:C_i_with_S0}, an issue must be overcome.
Indeed, due to their limited temporal and spatial span, our observations may not capture
all interactions between Iago and other pedestrians, in the non-static
scenarios. Therefore, the infection status $S_{j}^{0}$ at $t_0$ is difficult to evaluate. To circumvent the issue, we derive upper and lower bounds on $C_{i}$ that do not require
specific information about the $S_{j}^{0}$.

The upper bound is straightforward, because $S_{j}^{0}\leqslant1$
by definition, which implies that $C_{i}^{(\tau_{i})}\leqslant\bar{C}_{i}^{(\tau_{i})}\equiv\sum_{j\notin G_{i}}1-e^{-N_{ij}(t_{0},t_{0}+\tau_{i})}$.
This value can be converted into an hourly rate of new infections,
$\bar{C}_{i}^{(1\,\mathrm{h})}$, by multiplying it with $\frac{1\,\mathrm{h}}{\tau_{i}}$.

The lower bound requires more careful examination, wherein we will
set a maximal value to the number of people that have \emph{already
been }infected at $t_{0}$, under the hypothesis that only Iago is
contagious on the premises. Let $H_{i}=\left\{ j\ \mathrm{s.t.}\ j\notin G_{i}\,\mathrm{and}\,N_{ij}(t_{0},t_{0}+\tau_{i})>0\right\} $
be the group of unrelated pedestrians \emph{j} with whom Iago came
in contact while filmed, so that this number reads $\sum_{j\in H_{i}}I_{j}(t_{0})$.
If Iago and Juliette (\emph{j}) are unrelated, it is sensible to consider
that they have not been close to one another for more than a duration
$\delta\tau$ in the global scenario (we will set $\delta\tau=5\,\mathrm{min}$
for most scenarios), viz.,

\begin{equation}
\sum_{j\in H_{i}}I_{j}(t_{0})\leqslant\sum_{j\notin G_{i}}1-e^{-N_{ij}(t_{0}-\delta\tau,t_{0})}\leqslant\bar{C}_{i}^{(\delta\tau)},
\label{eq:SI_lower_bound1}
\end{equation}
where $\bar{C}_{i}^{(\delta\tau)}$ is our upper-bound estimate for
the number of infections during $\delta\tau$. Note that the foregoing
inequality manifestly holds if Iago and Juliette's interaction took
place in the interval $[t_{0}-\delta\tau,t_{0}]$, but it will also
hold should the interaction have occurred earlier, provided that our
video is representative. Indeed, under this proviso, Juliette is equally
likely to have already been infected at $t_{0}$ as any \emph{random}
pedestrian in Iago's vicinity in the interval $[t_{0}-\delta\tau,t_{0}]$
(note that these \emph{random} pedestrians are more numerous than
those observed in $[t_{0},t_{0}+\tau_{i}]$, because it was almost
always verified that $\tau_{i}<\delta\tau=5\,\mathrm{min}$). It follows from Eq.~\ref{eq:SI_lower_bound1}
that 

\[
\sum_{j\in H_{i}}S_{j}^{0}\geqslant\#H_{i}-\bar{C}_{i}^{(\Delta T)}.
\]

One is thus left with an optimization under constraints, wherein one
has to minimize (i.e., find a lower bound for) 

\begin{equation}
C_{i}^{(\tau_{i})}=\sum_{j\in H_{i}}S_{j}^{0}\cdot\left[1-e^{-N_{ij}(t_{0},t_{0}+\tau_{i})}\right]\label{eq:minimizing_eq}
\end{equation}
under the following constraints on the variables $S_{j}^{0}$,

\begin{equation*}
\forall j\in H_{i},  0 \leqslant S_{j}^{0} \leqslant1
\end{equation*}
\begin{equation*}
\sum_{j\in H_{i}}S_{j}^{0}  \geqslant  \#H_{i}-\bar{C}_{i}^{(\Delta T)}.
\end{equation*}
This optimization problem may for instance be solved using Lagrange's
multipliers. The minimum is reached at $(S_{j}^{0})_{j\in H_{i}}$
such that $S_{j}^{0}=1$ for all $j\in H_{i}$, \emph{except }
\begin{itemize}
\item the $n\equiv\left\lfloor \bar{C}_{i}^{(\Delta T)}\right\rfloor $
indices $j$ exhibiting the highest values $N_{ij}(t_{0},t_{0}+\tau_{i})$,
which are set to $S_{j}=0$, 
\item $j=n$, which is set to $S_{n}^{0}=1-\left(\bar{C}_{i}^{(\Delta T)}-\left\lfloor \bar{C}_{i}^{(\Delta T)}\right\rfloor \right)$.
\end{itemize}
In other words, this boils down to treating as infected the $n$
unrelated people who interacted most closely with Iago on the video.
The lower bound $\underline{C}_{i}^{(\tau_{i})}$ that we were seeking
is obtained by computing Eq.~\ref{eq:minimizing_eq} with the above
$(S_{j}^{0})_{j\in H_{i}}$, which concludes our search.

Finally, let us note that the foregoing reasoning can easily be extended
to a population with partial immunity, by replacing the initial conditions
$S_{j}^{0}=1-I_{j}(t_{0})$ with $S_{j}^{0}=(1-\alpha)\cdot\left[1-I_{j}(t_{0})\right]$,
where $\alpha$ is the fraction of immunized people.

\subsection*{Static scenarios}

In the 'static' scenarios, i.e., the streets caf\'es and the queue,
interactions are less changing. Therefore, we consider that the
interactions that are observed on the videos are prolonged over the
whole period of study, i.e., 1 hour to get an hourly rate. Accordingly,
the risk increments $N_{ij}(t_{0},t_{0}+\tau_{i})$ in Eq.~\ref{eq:evolution_Ij_2}
are simply multiplied by $\frac{1\,\mathrm{h}}{\tau_{i}}$, where
$\tau_{i}$ generally coincides with the duration of the video, and
the hourly rate of new infections is given by
\[
C_{i}^{(1\,\mathrm{h})}=\sum_{j}\left[1-e^{-\frac{1\,\mathrm{h}}{\tau_{i}}\cdot N_{ij}(t_{0},t_{0}+\tau_{i})}\right].
\]
A remark should be made here. The sum is not restricted to people
that are unrelated to Iago, contrary to the case of moving crowds,
and we assume that everybody apart from Iago is initially susceptible,
$S_{j}^{0}=1$. Indeed, infections at a caf\'e are counted starting
from the moment when people meet and sit together. Besides, on these
videos, we are not able to distinguish Iago's household members (who
may have been infected beforehand) from his other relatives or friends,
so everybody is assumed initially susceptible.

\section{\label{sec:SI_dynamic_model}Dynamic model of viral transmission}

The main text of this paper focused on a family of \emph{ad hoc} models of viral transmission $\nu(t)$ that overlook the propagation dynamics of respiratory droplets. In this supplemental section, we generalize our methodology by considering \emph{dynamical} transmission models derived from microscopic fluid dynamics computations of droplet propagation. This will allow us to ascertain that our main findings remain valid with more realistic transmission models and, in particular, that the differences in the estimated infection risks between scenarios are not an artifact due to our simple models, but arise from intrinsic differences between the scenarios. Admittedly, the more realistic models used in this section also suffer from some inaccuracies, by accounting neither for the wind nor for differences in walking speeds, in activity (speech, cough), and so on; a more exhaustive study including these effects is differed to a subsequent publication. Still, they tend to bolster the robustness of our main results.

The propagation of droplets emitted during a pedestrian's breathing cycle is simulated with the CFD (Computational Fluid Dynamics) software YALES2 \cite{Moureau:2011design} (https://www.coria-cfd.fr/index.php/YALES2), already used in the context of the transport of respiratory droplets \cite{abkarian2020speech,yang2020towards}. We first compute the air flow field around a standing person ($v=0\,\mathrm{m}\cdot\mathrm{s}^{-1}$) and the flow field around a pedestrian walking at speed $v=1\,\mathrm{m}\cdot\mathrm{s}^{-1}$, separately. In the latter case, the simulation is performed in the moving frame of the walker, so that the difference of speed between the pedestrian and the ambient air is mimicked by a uniform airflow at  $v=1\,\mathrm{m}\cdot\mathrm{s}^{-1}$ opposite to  the walking velocity (the person thus remains static and body motion is not accounted for). The pedestrian is breathing out through the mouth with a breathing period of 3~s, with 1.5~s of exhaling and inhaling times, and an exhaled flow rate of 20~L/min, which is representative of mild exercise. The mouth region delimited by the lips has an area of 5~cm$^2$, approximately. Uniform flow is imposed at exhalation, with a flow along the walking direction. The boundaries of the domain are typically 2~m away from the walker; the grid is fully tetrahedral, with a mesh size as low as 5~mm in the vicinity of the head  and a uniform mesh size of 8~mm in the rest of the region of interest where droplets flow. This is dealt with automatic mesh refinement  \cite{Benard2016mesh} which adapts the mesh size along the course of the calculation.

\begin{figure*}[h]
\begin{centering}
\includegraphics[width=1\textwidth]{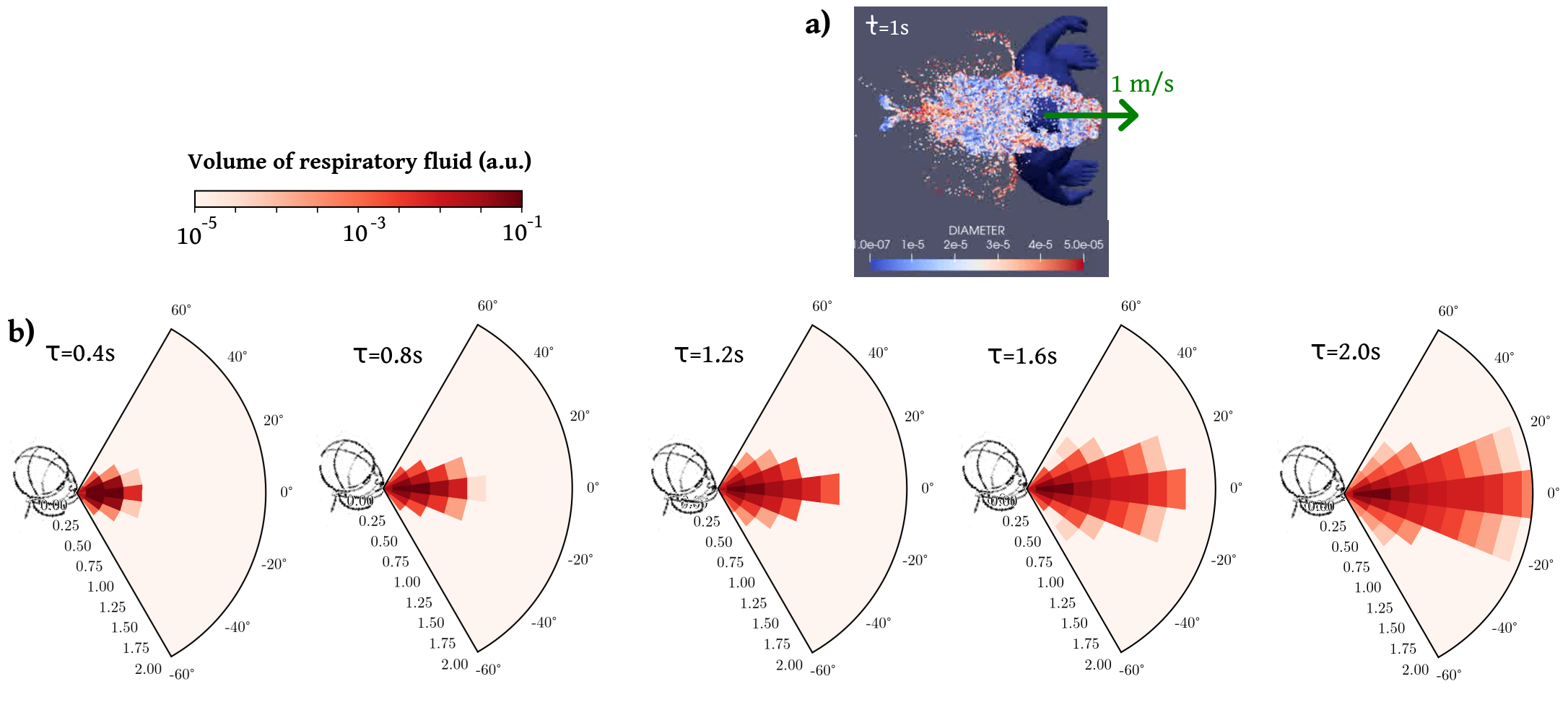}
\par\end{centering}
\caption{\label{fig:SI_fig_dynamic_trans}Dynamic model of viral transmission. \textbf{(a)} Top view of the propagation of respiratory droplets around a pedestrian walking at 1 $\mathrm{m}\cdot \mathrm{s}^{-1}$, simulated with CFD, in the walker's co-moving frame. A large number of droplets is visible. In the range of diameter between 0.1 and 50 microns, an arbitrarily large number ($\sim$60,000) of droplets are injected per breath; this number exceeds actual emissions by far but it will not impact the calculated transmission rates, which are normalised by $T_0$. The quantity of virus in droplets of a given size is then suitably rescaled to match the target exhaled size distribution \cite{johnson2011modality}. \textbf{(b)} Spatio-temporal representation of the coarse-grained disease transmission rate, in the laboratory frame, as a function of the propagation delay $\tau$. Note the logarithmic scale. }
\end{figure*}

Then, we simulate the propagation of respiratory droplets of various diameters in these flow fields, as illustrated in the top view of Fig.~\ref{fig:SI_fig_dynamic_trans}(a). As the issue of droplet evaporation is still debated and its impact may have been overestimated in the past \cite{chong2021extended}, we chose not to account for evaporation here. As a first approximation, we thus performed simulations at constant temperature, so that incompressible Navier-Stokes equations are solved, with a uniform kinematic viscosity of $1.5\cdot 10^{-5}\, \mathrm{m}^2\cdot\mathrm{s}^{-1}$. Buoyancy effects are thus neglected, but they are not expected to dominate in this configuration where rapid mixing occurs between the exhaled air and the ambient air. We use Large-Eddy Simulations with the so-called sigma subgrid model \cite{Nicoud2011using} and the numerical method is identical to the one recently used by Abkarian et al.  \cite{abkarian2020speech} for simulations of the flow generated by speech. The simulation is run for several cycles (e.g., 7 cycles, i.e., 21~s, for the walking pedestrian): the flow is established during the first 3 cycles and only the subsequent cycles are used to compute the coarse-grained transmission rates.

Another issue is also prickly, that of the sizes of exhaled droplets. Widely different distributions of sizes can be found in the literature, leading to widely different travel ranges \cite{rosti2020fluid}. Here, we arbitrarily selected Johnson et al.'s \cite{johnson2011modality}
data and focused on the breathing mode, for which the distribution of droplet diameters $D$ is log-normal and its cumulative function $P_s$ obeys
\begin{equation}
dP_s(D) \propto e^{- \frac{\mathrm{ln}^2(D/\bar{D})}{2  \mathrm{ln}^2\sigma} }\,d\mathrm{ln}D, 
\end{equation}
where $\bar{D}\simeq 0.8\,\mathrm{\mu m}$ and $\sigma\simeq 1.3$. 

Provided that exhaled droplets have an equal concentration of viral copies, the disease transmission rate $\nu(t_e,t_e+\tau)$ due to droplets shed at $t_e$ and inhaled at $t_e+\tau$ is proportional to the mass concentration of droplets at the receiver's position at $t_r$, relative to the emitter's position at $t_e$, within a $40\,\mathrm{cm}$-thick horizontal slice centered on the emitter's mouth. The spatio-temporal evolution of this rate is symmetrized and coarse-grained by binning the data into $\tau$ of duration 0.2~s and polar cells of radial length 20~cm. The question of the number of emitted droplets and that of the viral load are eluded by setting a characteristic infection time $T_0$ of 10~min for someone standing face-to-face with a static emitter at a distance $r_c=50\,\mathrm{cm}$. Finally, the effect of the inhaler's head orientation $\theta^R$ is accounted for by multiplying this rate by a factor $\mathrm{exp} (1-\frac{|\theta^R|}{\pi / 3})$, as in our \emph{standard} models coupled with $f_3$. This dependence is not established on the basis of CFD simulations; doing so would entail several additional complications.\\
The resulting coarse-grained rates are found to display a fairly similar spatio-temporal pattern for the standing emitter and for the walking one, shown in Fig.~\ref{fig:SI_fig_dynamic_trans}(b) in the laboratory frame, except that in the former case the propagation is slower and shorter-ranged (in the laboratory frame) whereas the risk induced by the walker is more diffuse in space. \\

\begin{figure}[ht]
\begin{centering}
\includegraphics[width=0.45\textwidth]{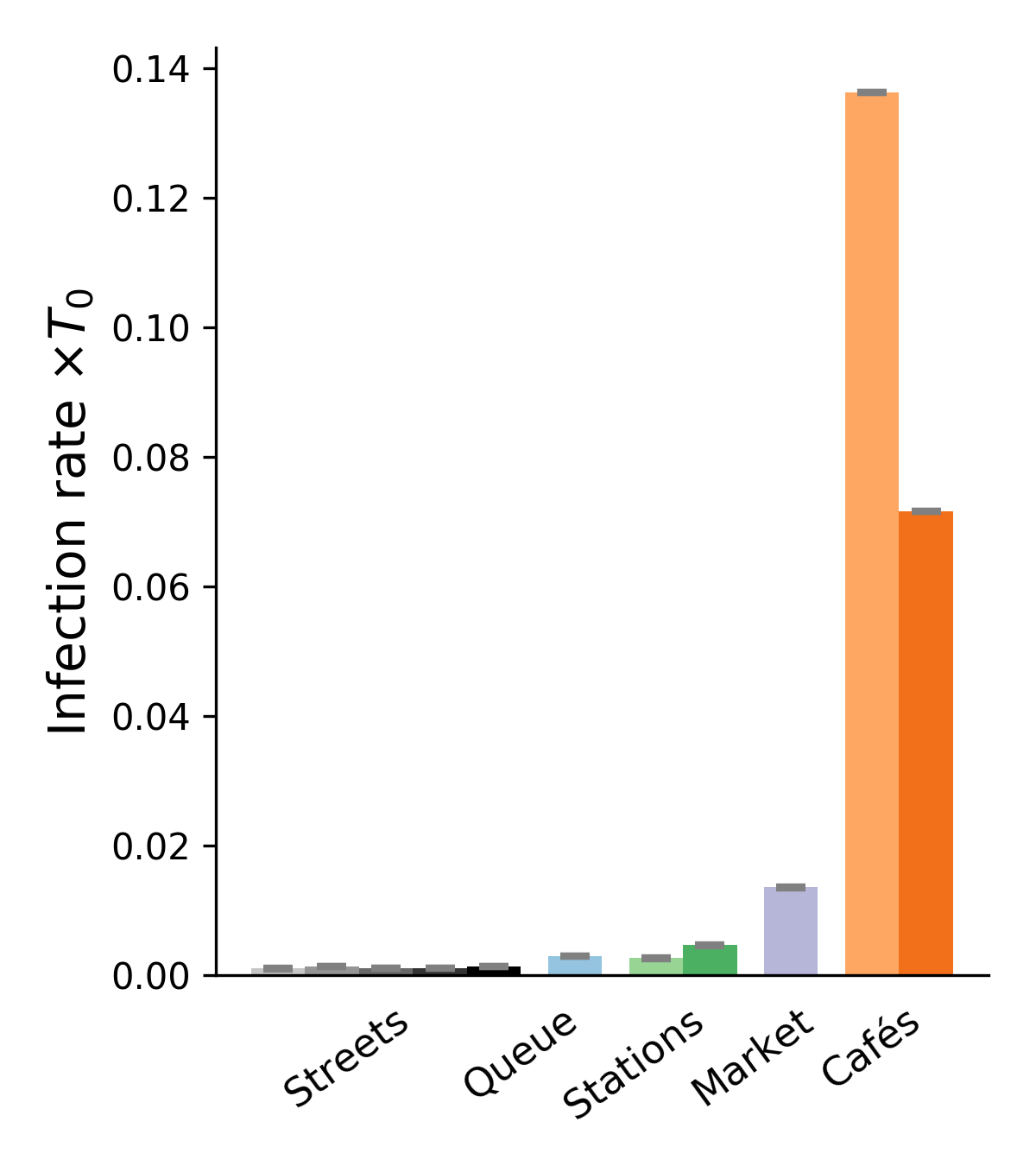}~
\includegraphics[width=0.45\textwidth]{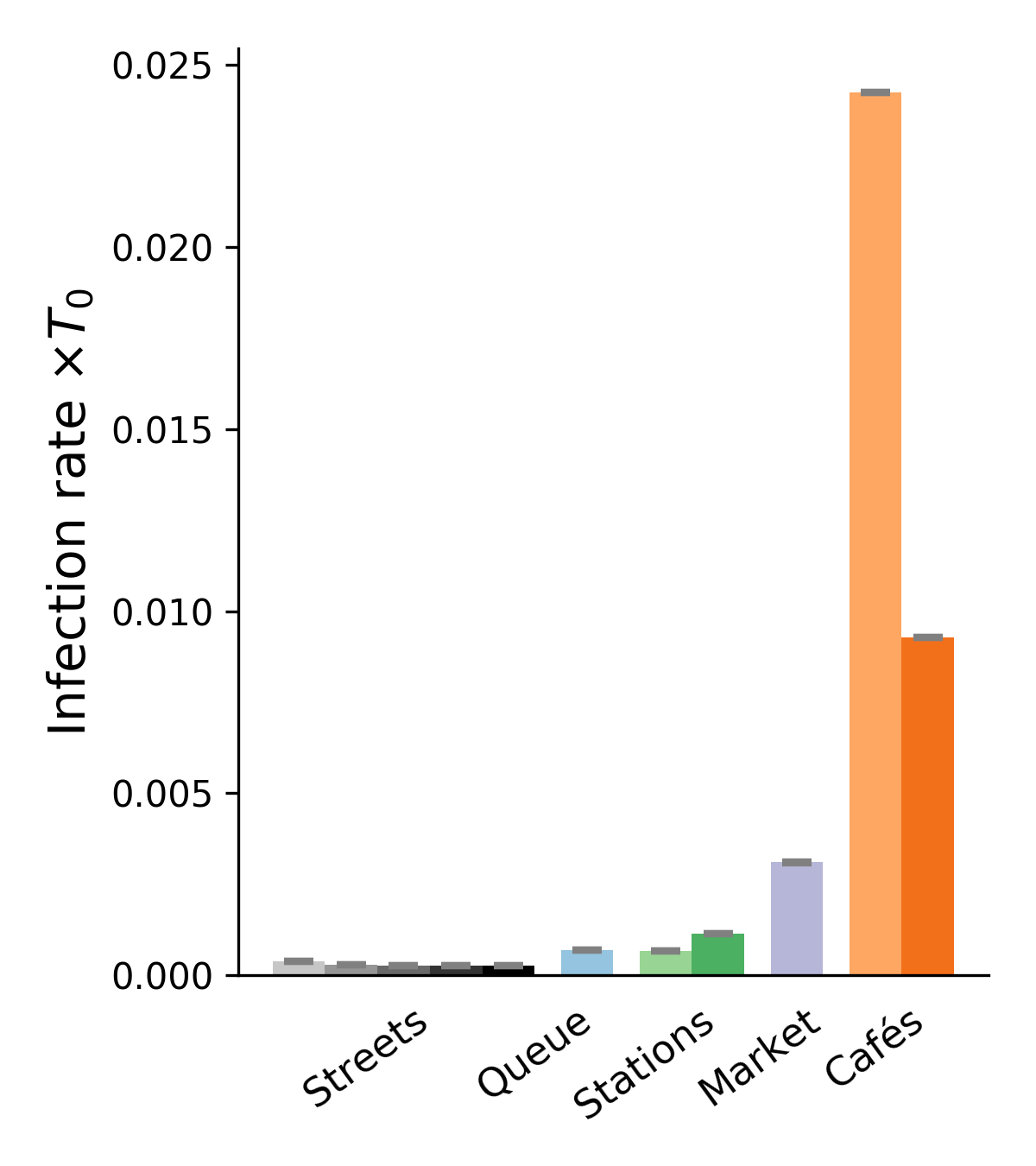}
\par\end{centering}
\caption{\label{fig:SI_fig_histo_dynamic}Number of new infections over a time interval of $T_0$ estimated with a dynamic transmission model, derived from CFD simulations of the propagation of droplets exhaled by a static pedestrian (\emph{left}) or a pedestrian walking at $v=1\,\mathrm{m\cdot s^{-1}}$ (\emph{right}). The characteristic infection time $T_0$ was set to 10~min for the computation. }
\end{figure}

Coupling the dynamic models\footnote{ For the street caf\'es and the queuing scenario, since most people are static, we simplified the model by integrating the dynamic transmission rates over the time delay $\tau$, viz., $\nu(t_e)=\int_0^{\infty} d\tau\,\nu(t_e,t_e+\tau)$.} to our field data yields the ranking of scenarios by the risks they present that is displayed in Fig.~\ref{fig:SI_fig_histo_dynamic}. Here, as a first approach, we have made the choice to apply 
the same model to all scenarios in each panel of Fig.~\ref{fig:SI_fig_histo_dynamic} [be it the `standing' one on the left panel or the `moving' one on the right panel], irrespective of the individual walking velocities. We agree that this choice is questionable, but at present we are unable to assign a speed-adjusted emission model to each pedestrian and all moving scenarios involve people walking at different speeds and a significant fraction of nearly static ones. Albeit questionable, our choice has the merit of not resting on an arbitrary, scenario-specific mingling of the transmission models.\\

The main conclusion to be drawn from Fig.~\ref{fig:SI_fig_histo_dynamic}  is that the hierarchy of risks (dominated by street caf\'es\footnote{Incidentally, the intriguing discrepancy between the two street caf\'e scenarios also owes to this short transmission range, whereby the crowd's configuration is probed at fine length scales; a similar tendency could already be observed with our optimistic models.}, with fairly busy streets very far behind) is identical to that found with our \emph{ad hoc} models. Besides, we have checked that varying the characteristic time $T_0$ (within reasonable bounds) does not alter the results much: Increasing it from 10~min to 20~min, for instance, more or less uniformly halves the predicted infection rates. These observations invite us to take the predicted rate values with a pinch of salt, given their sensitivity to a variety of details, but above all they further bolster the relative scenario rankings put forward in the paper. 

\begin{figure}[ht]
\begin{centering}
\includegraphics[width=0.45\textwidth]{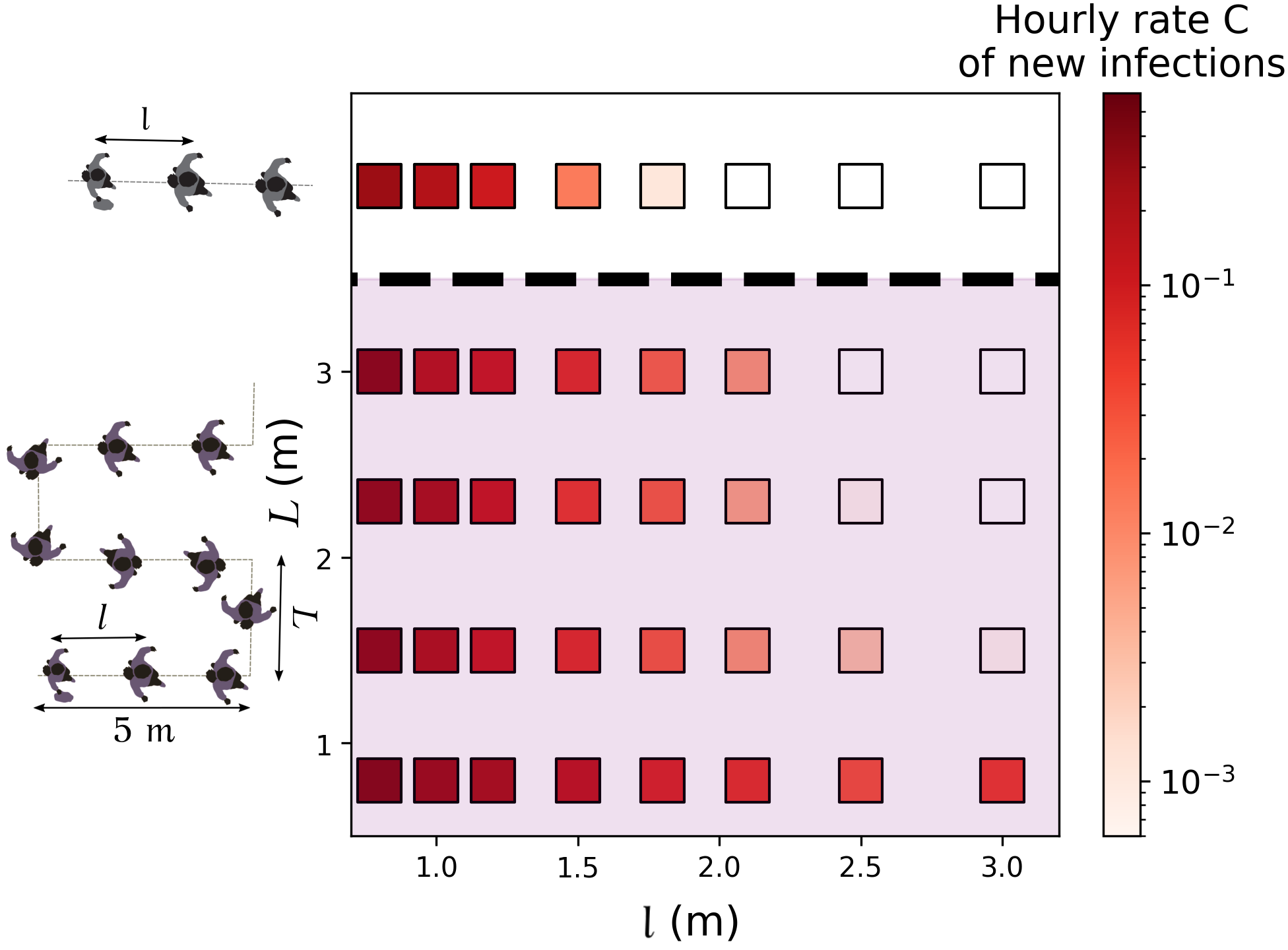}
\par\end{centering}
\caption{\label{fig:SI_fig_dynamic_queuing_rates}Hourly rate of new infections in a queue, as a function of the queuing geometry. The characteristic infection time $T_0$ is set to 10~min.}
\end{figure}

Finally, we revisited our predictions regarding the effect of the queuing geometry in the light of these more realistic models. For this purpose, the rate of new infections in a queue were computed afresh, depending on its geometry and using the CFD transmission model for standing people ($v=0\,\mathrm{m\cdot s^{-1}}$), integrated over time delays, as the queuing pedestrians are mostly static. The results are presented in Fig.~\ref{fig:SI_fig_dynamic_queuing_rates} and, once again, are qualitatively very similar to those obtained previously, even though quantitatively smaller rates are found.

\section{Additional figures}

\begin{figure}[ht]
\begin{centering}
\includegraphics[width=1\textwidth]{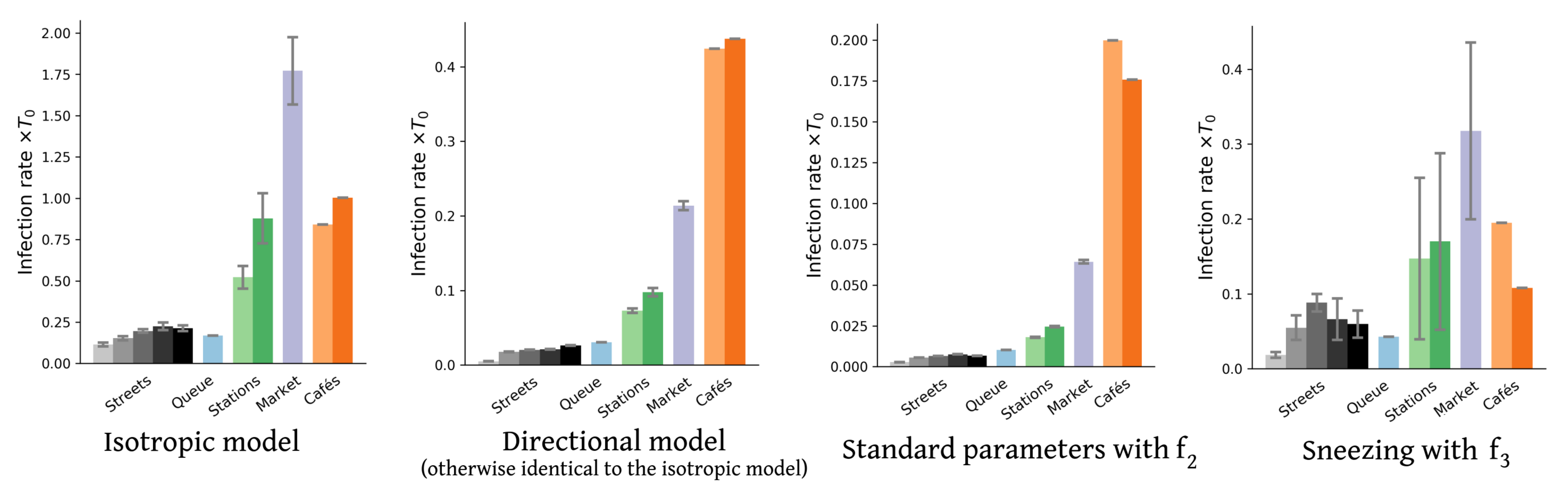}
\par\end{centering}
\caption{\label{fig:SI_Ranking_of_scenarios_bis}Hourly rates of new infections multiplied by $T_0$  in the scenarios under study, estimated with four transmission
parameter sets. The isotropic model and the directional one ($\theta^E_0=\frac{\pi}{6}$, $\theta^R_0=\frac{\pi}{3}$) share the same parameters $T_0=15\,\mathrm{min}$ and $r_0=1\,\mathrm{m}$. For the standard model and the model describing uncovered sneezes, the characteristic times $T_0$ were set to $10\,\mathrm{min}$ and $1.7\,\mathrm{min}$ (see Table~1 of the main text), respectively, but in reality these will be sensitive to a variety of details. The error bars span the interval between the estimated lower
bound $\underline{C}^{(T_0)}$ and upper bound $\bar{C}^{(T_0)}$,
while the filled bars represent $\frac{1}{2}\left(\underline{C}^{(T_0)}+\bar{C}^{(T_0)}\right)$. Refer to Table~S1 for details about the investigated scenarios.}
\end{figure}

\begin{figure}[ht]
\begin{centering}
\includegraphics[width=1\textwidth]{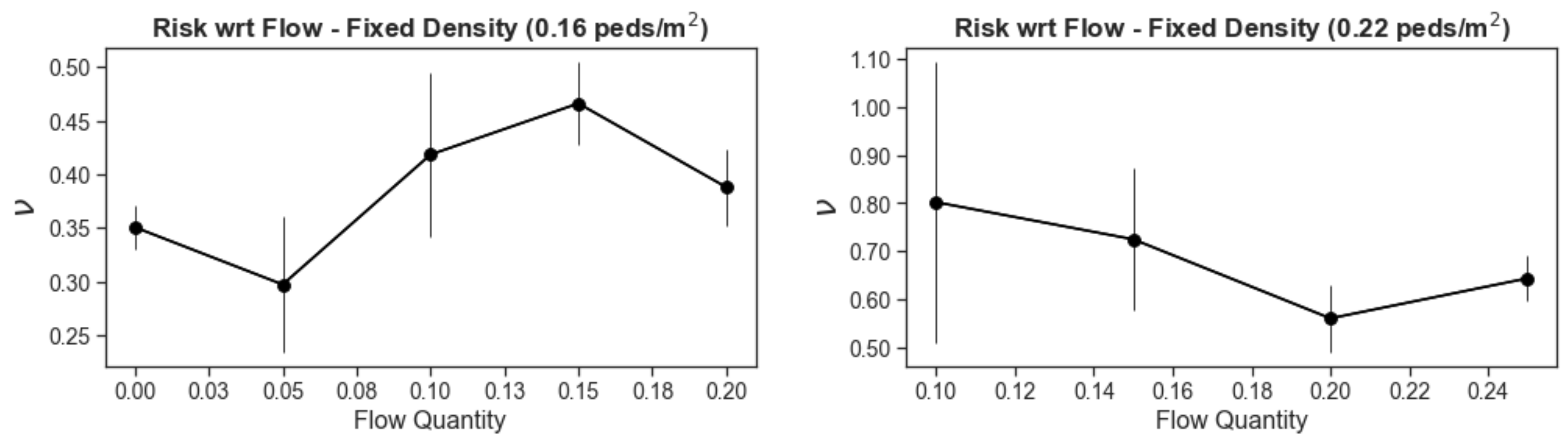}
\par\end{centering}
\caption{\label{fig:SI_risk_vs_flow}Variations of the global transmission rate $\nu$ with the total flow rate in a street in the Old Town of Lyon (\emph{left}) and along an outdoor market alley (\emph{right}), at fixed pedestrian density, with \emph{ModOpt}$_3$. }
\end{figure}

\begin{figure}
\begin{centering}
\includegraphics[width=0.4\textwidth]{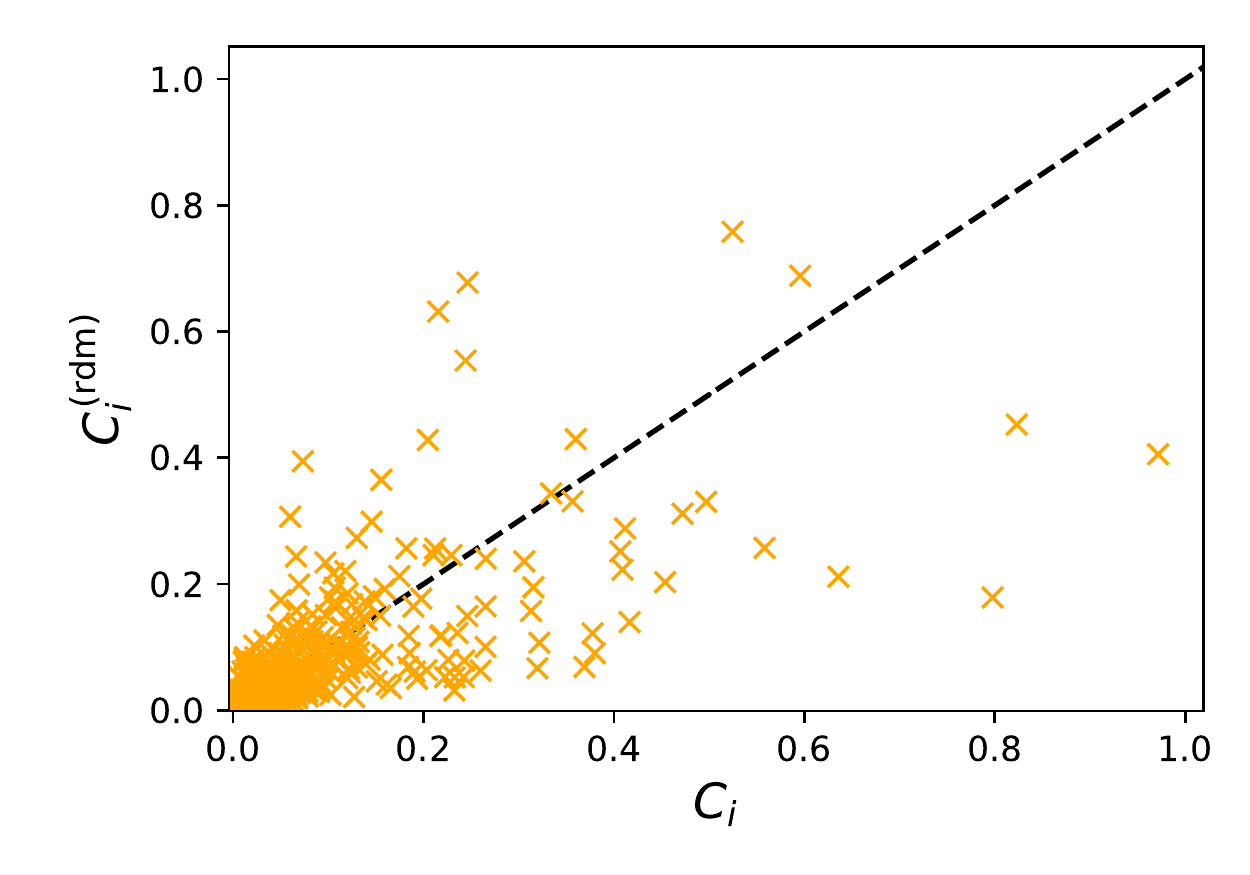}
\includegraphics[width=0.4\textwidth]{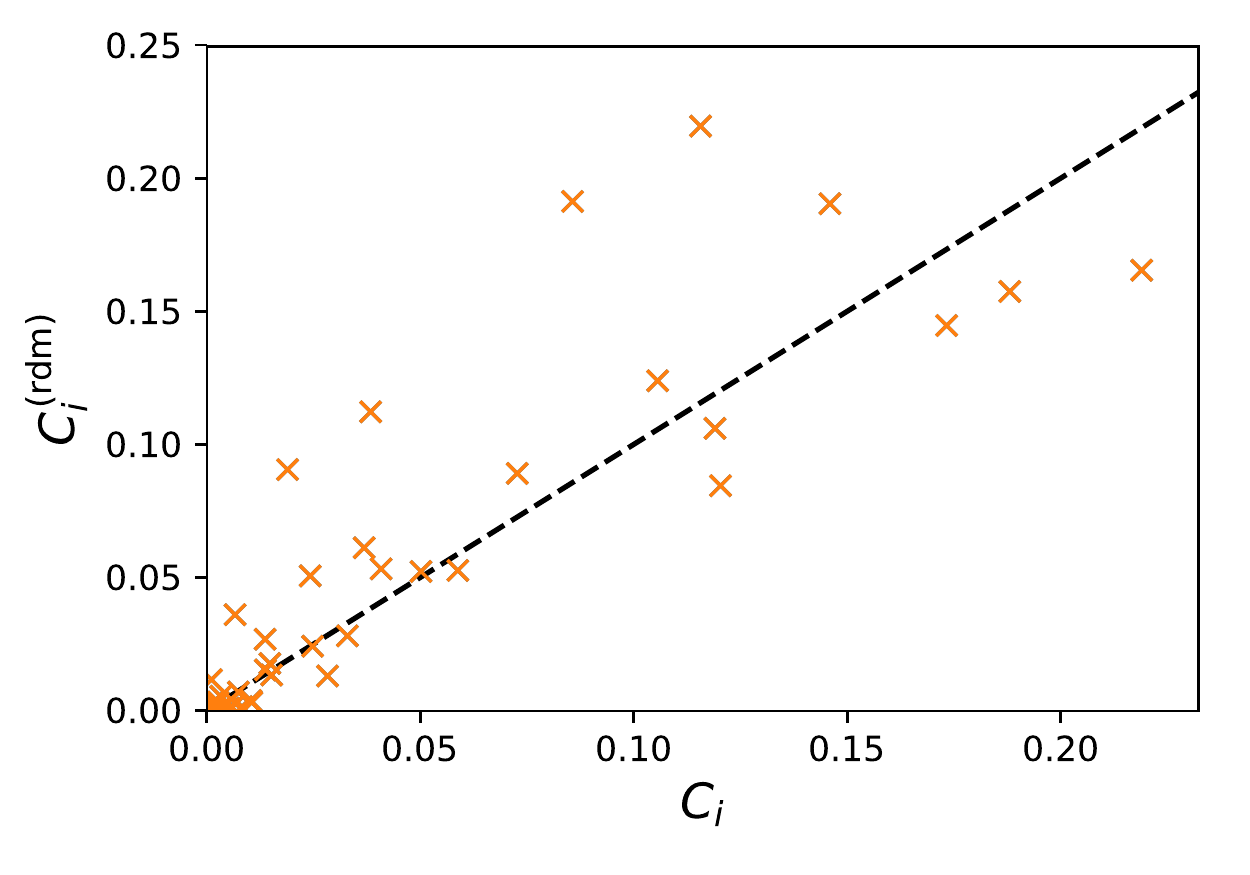}
\par\end{centering}
\caption{\label{fig:SI_Cbari_vs_randomOr} Correspondence between the individual-based hourly rates of new infections $\bar{C}_i$ based on the \emph{bona fide} head orientations (upper bound) and the values $\bar{C}_i^{(rdm)}$ reconstructed under the assumption of random head orientations (see the main text), on the plaza in front of the Perrache train station (\emph{left}) and on the Rhône riverbank (\emph{right}), with \emph{ModOpt}$_3$. }
\end{figure}

\newpage

\bibliographystyle{elsarticle-num}
\bibliography{Covid}

\end{document}